\documentstyle[prd,aps,amsmath,latexsym,amssymb,floats]{revtex}
\renewcommand{\theequation}{\arabic{section}.\arabic{equation}}

\def \slas{\kern -6.2pt /}
\def \sla{\kern -5.4pt /}
\def \sl{\kern -4.0pt /}
\def \Cslas{\kern -6.8pt /}
\def \Dslas{\kern -7.4pt /}
\def \slass{\kern -7.4pt /}
\def \OO{{\cal O}}

\def \ii{{\mathrm{i}}}
\def \d{{\mathrm{d}}}

\def \lcd{\tilde{\partial}}
\def \pd{\partial}
\def \e{{\mathrm{e}}}

\def \lcx{\tilde{x}}
\def \tl#1{\overset{\kern 2pt\circ}{#1}}
\def \tll#1{\overset{\kern -1pt\circ}{#1}}
\def \TL#1{\overset{\kern -28pt \circ}{#1}}
\def \TLL#1{\overset{\kern -7pt \circ}{#1}}

\begin{document}
\draft
\title{Parton distribution functions from nonlocal light-cone operators 
with definite twist}
\author{Bodo Geyer\thanks{E-mail: geyer@itp.uni-leipzig.de}
	and 
	Markus Lazar\thanks{E-mail: lazar@itp.uni-leipzig.de}}
\address{Center for Theoretical Studies
	 and Institute of Theoretical Physics,\\
	Leipzig University, 
	Augustusplatz~10, D-04109~Leipzig, Germany} 
\date{\today}    
	
\maketitle

\hspace*{2.0cm}

\begin{abstract}
We introduce the chiral-even and chiral-odd quark distributions
as forward matrix elements of related bilocal quark operators
with well-defined (geometric) twist. Thereby, we achieve
a Lorentz invariant classification of these distributions
which differ from the conventional ones by explicitly taking
into account the necessary trace terms. The relations between both kinds of
distribution functions are given and the mismatch between their different 
definition of twist is discussed. Wandzura-Wilczek--like relations 
between the conventional distributions (based on dynamical twist) are 
derived by means of geometric twist distribution functions.
\end{abstract}
\hspace*{0.5cm}

\section{Introduction}
The scattering amplitudes of light-cone dominated hadronic processes --
according to the factorization hypothesis -- are usually represented
by the convolution of the hard (process-dependent) scattering amplitude 
of the partons with appropriate soft (process-independent) parton distribution
amplitudes. The different phenomenological distribution amplitudes, 
e.g.,~parton distribution functions in deep inelastic scattering and 
in Drell-Yan processes as well as hadronic wave functions are given as 
(non-forward) matrix elements of some nonlocal light-cone (LC) operators.
Thereby, one and the same LC operator is related to different hadronic
processes. Moreover, the classification of these distribution 
functions suffers from the fact, that these operators by its twist 
decomposition also contribute to different distribution functions. 
Therefore, it is necessary to disentangle these various 
contributions of the original operator and to specify the contributions
of the trace terms. 

Quite recently, a general group theoretical procedure has been introduced to 
decompose these nonlocal LC operators~\cite{glr99b,gl00a} 
into operators of definite twist. This procedure is based on the notion of 
{\it geometric} twist = mass dimension -- (Lorentz) spin, $\tau=d-j$, 
originally introduced by Gross and Treiman~\cite{Gross}. On the other hand,
Jaffe and Ji~\cite{Jaf92} proposed the notion of {\it dynamical} 
twist ($t$) by counting powers $Q^{2-t}$ which is directly related to 
the power by which the corresponding distributions contribute to the 
scattering amplitudes. They found nine independent forward quark distribution 
amplitudes. Recently, Ball {\it et al.} have used this pattern for the 
consideration of higher twist distribution
amplitudes of vector mesons in QCD~\cite{ball98,ball99}. 

However, that dynamical notion of twist is only defined for 
the {\it matrix elements} of operators, it is not Lorentz invariant and also 
not simply related to the contributions of (higher) geometric twist:
Both definitions of twist do not coincide at higher orders.
Thus, the definition of dynamical twist has a mismatch with the conventional 
definition of geometrical twist. This mismatch gives rise to relations of  
Wandzura-Wilczek type which show that dynamical twist distributions contain 
various parts of
different geometric twist. With the growing accuracy of data 
for QCD tests, the higher twist contributions become relevant. Therefore,
it seems to be quite urgent to consider these contributions on a rigorous 
quantum field theoretical basis. 
Jaffe {\it et al.} have already pointed out that the geometric twist is 
the most convenient one to discuss higher twist effects on a 
reliable basis~\cite{Jaf92,JS82,jaffe83}. In addition, it should be
emphasized that the LC operators of definite geometric twist are to be
preferred also from the point of view of renormalization theory, because they 
might be assumed to have much simpler renormalization properties then 
those used in the definition of the conventional structure functions.

The aim of this paper is to present the {\em forward quark distribution 
functions} which are related to the nonlocal LC-operators of different
geometric twist. In that framework, it is possible to investigate 
in an unique manner the contributions resulting from the traces of the 
operators having well-defined twist. By the way, the classification of the 
various distribution functions appears to be quite straightforward.

The paper is organized as follows. Sect.~2 is mainly introductory.
It collects the necessary definitions and explains the basic ideas.
Sect.~3 contains a detailed study of chiral-odd as well as chiral-even 
distribution functions and of the forward matrix elements with definite 
geometric twist. Whereas the preceeding sections contain the group
theoretical aspects the next sections are devoted to the phenomenological
aspects. The interrelation between the conventional and the new structure
functions of dynamical and geometric twist, respectively, is given in Sect.~4. 
Thereby, it is proven that both types of structure functions could be used
for the parametrization of experimental data because they 
can be expressed uniquely through each other. However, the virtue
of the new structure functions is their well defined twist. This allows
also to disentangle the genuine twist content of the conventional distribution
functions. This is used in Sect.~5 to derive Wandzura-Wilczek--like 
relations between the conventional distribution functions. Most of them are 
new and might be of phenomenological interest. Finally, Sect.~6 contains a 
summary and conclusions.  App.~A collects some useful 
formulae about the interior derivative on the light-cone. Additionally, 
we rewrite our operators of definite geometric twist by means of 
this derivative. App.~B clarifies some transformation properties of
nonlocal operators with definite geometric twist.
 
\section{Twist decomposition of bilinear quark operators}
\setcounter{equation}{0}

At first, we shortly review the systematic procedure of group theoretical 
twist decomposition for local and nonlocal operators~\cite{glr99b,gl00a}.
It is based on the unique spin decomposition with respect to the 
Lorentz group $SO(3,1)$ of the considered operators.  
The consecutive steps which are used to decompose a bilocal operator,
for example, $O_{\alpha}(\kappa_1 x,\kappa_2 x)=
\bar{\psi}(\kappa_1 x)\gamma^5\gamma_{\alpha}
U(\kappa_1 x,\kappa_2 x)\psi(\kappa_2 x)$, into operators
of definite geometric twist are the following:\\
(1) {\it Taylor expansion} of the nonlocal tensor operator for 
{\em arbitrary} values of $x$  
into an infinite tower of {\em local} tensor operators having 
rank $n$ and dimension $d$;\\
(2) {\it Decomposition} of these local tensor operators into
{\it irreducible} ones -- with respect to the Lorentz group -- by means of 
the standard method characterizing the symmetry class by Young patterns
$[m]= (m_1,m_2, \ldots,m_r)$ of $r$ rows;\\
(3) {\it Resummation} of the irreducible local tensor operators belonging 
to the same symmetry class and having the same twist $\tau$
to a nonlocal tensor operator with definite twist;\\
(4) {\it Projection} of the nonlocal twist operator onto the light--cone,
$x \rightarrow \lcx$ with $\lcx^2=0$, thereby reducing the infinite twist 
decomposition to a finite one.

In~\cite{glr99b} we applied that procedure to the following bilocal
quark operators on the light-cone having (axial) vector as well as 
skew tensor structure:
\begin{align}
\label{O_ent}
O_{5\alpha}(\kappa_1\lcx,\kappa_2\lcx)&=
\bar{\psi}(\kappa_1\lcx)
\gamma^5
\gamma_{\alpha}
U(\kappa_1\lcx,\kappa_2\lcx)\psi(\kappa_2\lcx),\\
\label{M_ent}
M_{5[\alpha\beta]}(\kappa_1\lcx,\kappa_2\lcx)
&=
\bar{\psi}(\kappa_1\lcx)\ii
\gamma^5\sigma_{\alpha\beta}
U(\kappa_1\lcx,\kappa_2\lcx)\psi(\kappa_2\lcx),
\end{align}
with the path ordered gauge factor along the straight line connecting the 
points $\kappa_1\lcx$ and $\kappa_2\lcx$:
\begin{equation}
\label{phase}
U(\kappa_1 \lcx, \kappa_2 \lcx)
 =  P \exp\left\{\ii g
\int_{\kappa_1}^{\kappa_2} \d w \,\lcx^\mu A_\mu(w \lcx)
\right\},
\end{equation}
and to the corresponding operators without $\gamma_5$ and
the derived scalar and vector operators, 
$O_{(5)}(\kappa_1\lcx,\kappa_2\lcx) = x^\alpha
O_{(5)\alpha}(\kappa_1\lcx,\kappa_2\lcx)$ and 
$M_{(5)\alpha}(\kappa_1\lcx,\kappa_2\lcx)= x^\beta
M_{(5)[\alpha\beta]}(\kappa_1\lcx,\kappa_2\lcx)$, respectively.

The resulting decomposition for the vector and skew tensor operators
-- which is independent of the presence or absence of $\gamma_5$ --
after a straightforward calculation is:
\begin{align}
\label{O_tw_nl}
O_{\alpha}(\kappa_1\lcx,\kappa_2\lcx)&=
 O^{\mathrm{tw2}}_{\alpha}(\kappa_1\lcx,\kappa_2\lcx)
+O^{\mathrm{tw3}}_{\alpha}(\kappa_1\lcx,\kappa_2\lcx)
+O^{\mathrm{tw4}}_{\alpha}(\kappa_1\lcx,\kappa_2\lcx)
,\\
\label{M_tw_nl}
M_{[\alpha\beta]}(\kappa_1\lcx,\kappa_2\lcx)&=
M^{\mathrm{tw2}}_{[\alpha\beta]}(\kappa_1\lcx,\kappa_2\lcx)
+M^{\mathrm{tw3}}_{[\alpha\beta]}(\kappa_1\lcx,\kappa_2\lcx)
+M^{\mathrm{tw4}}_{[\alpha\beta]}(\kappa_1\lcx,\kappa_2\lcx)
,\\
\label{M_vec}
M_{\alpha}(\kappa_1\lcx,\kappa_2\lcx)&
=M^{\mathrm{tw2}}_{\alpha}(\kappa_1\lcx,\kappa_2\lcx)
+M^{\mathrm{tw3}}_{\alpha}(\kappa_1\lcx,\kappa_2\lcx).
\end{align}
with (using the convention
$a_{[\mu}b_{\nu]} \equiv (1/2) (a_{\mu}b_{\nu} - a_{\nu}b_{\mu})$) 
\begin{align}
\label{O2sca}
O^{\rm tw2}(\kappa_1\lcx,\kappa_2\lcx)
&=
\lcx^\mu O_\mu(\kappa_1\lcx,\kappa_2\lcx)
= \bar{\psi}(\kappa_1\lcx)(\gamma\lcx)
U(\kappa_1\lcx, \kappa_2\lcx) \psi(\kappa_2\lcx)
\\
\label{O2vec}
O^{\mathrm{tw2}}_{\alpha}(\kappa_1\lcx,\kappa_2\lcx)
&=
\int_{0}^{1} \d\lambda
\Big(\pd_\alpha +
\hbox{\Large$\frac{1}{2}$}(\ln\lambda)\,x_\alpha\square\Big)
x^\mu
O_\mu(\kappa_1\lambda x, \kappa_2\lambda x)
\big|_{x=\tilde{x}}
\\
\label{O3vec}
O^{\mathrm{tw3}}_{\alpha}
(\kappa_1\lcx,\kappa_2\lcx)
&=
\int_{0}^{1}\d\lambda
\Big(\delta_\alpha^\mu(x\pd)-
x^\mu\pd_\alpha-(1+2\ln\lambda)x_\alpha\pd^\mu
-(\ln\lambda)\, x_\alpha x^\mu\square
\Big) O_\mu(\kappa_1\lambda x, \kappa_2\lambda x)
\big|_{x=\tilde{x}}
\\
\label{O4vec}
O^{\mathrm{tw4}}_{\alpha}
(\kappa_1\lcx,\kappa_2\lcx)
&=\lcx_\alpha
\int_{0}^{1}\d\lambda\Big(
(1+\ln\lambda)\pd^\mu+
\hbox{\Large$\frac{1}{2}$}(\ln\lambda)\,x^\mu \square
\Big)
O_\mu(\kappa_1\lambda x, \kappa_2\lambda x)\big|_{x=\tilde{x}}
\\
\label{M_tw2_ten}
M^{\mathrm{tw2}}_{[\alpha\beta]}(\kappa_1\lcx,\kappa_2\lcx)
&=
 \int_{0}^{1}\d\lambda\Big\{2\lambda\,
\pd_{[\beta}\delta_{\alpha]}^\mu
-(1-\lambda)\big(2x_{[\alpha}\pd_{\beta]}\pd^\mu
-x_{[\alpha}\delta_{\beta]}^\mu\square\big)\Big\}
x^\nu M_{[\mu\nu]}(\kappa_1\lambda x, \kappa_2\lambda x)
\big|_{x=\tilde{x}}
\\
\label{M_tw3_ten}
M^{\rm tw3}_{[\alpha\beta]}(\kappa_1\lcx,\kappa_2\lcx)
&=
\int_{0}^{1}\d\lambda
\Big\{\lambda\big((x\pd)\delta_{[\beta}^\nu
- 2x^\nu\pd_{[\beta}\big)\delta_{\alpha]}^\mu
+\hbox{\Large$\frac{1-\lambda^2}{\lambda}$}
\Big(x_{[\alpha}\big(
\delta_{\beta]}^{[\mu} (x\pd)
-
x^{[\mu}\pd_{\beta]}
\big)\pd^{\nu]}
-x_{[\alpha}\delta_{\beta]}^{[\mu}x^{\nu]}\square
\nonumber
\\
&\qquad
-x_{[\alpha}\pd_{\beta]}x^{[\mu}\pd^{\nu]}
\Big)\Big\}
M_{[\mu\nu]}(\kappa_1\lambda x, \kappa_2\lambda x)\big|_{x=\lcx}
\\
\label{M_tw4_ten}
M^{\rm tw4}_{[\alpha\beta]}(\kappa_1\lcx,\kappa_2\lcx)
&=
\int_{0}^{1}{\d\lambda}
\hbox{\Large$\frac{1-\lambda}{\lambda}$}
\Big\{x_{[\alpha}\delta_{\beta]}^{[\mu}
x^{\nu]}\square
-2x_{[\alpha}\big(
\delta_{\beta]}^{[\mu} (x\pd)
-
x^{[\mu}\pd_{\beta]}
\big)\pd^{\nu]}
\Big\}
M_{[\mu\nu]}(\kappa_1\lambda x, \kappa_2\lambda x)\big|_{x=\lcx}
\\
\label{M_tw2_vec}
M^{\mathrm{tw2}}_{\alpha}(\kappa_1\lcx,\kappa_2\lcx)
&=
M_\alpha(\kappa_1\lcx, \kappa_2\lcx)
-\lcx_\alpha \pd^\mu
\int_0^1\d\lambda\,\lambda\,
M_{\mu}(\kappa_1\lambda x, \kappa_2\lambda x)
\big|_{x=\tilde{x}}
\\
\label{M_tw3_vec}
M^{\mathrm{tw3}}_{\alpha}(\kappa_1\lcx,\kappa_2\lcx)
&=
\lcx_\alpha\pd^\mu
\int_0^1\d\lambda\,\lambda
M_\mu(\kappa_1\lambda x, \kappa_2\lambda x)\big|_{x=\tilde{x}}.
\end{align}

Obviously, the trace terms are those being
multiplied by $\lcx^\alpha$ (or $\lcx^\beta$).

Let us remark that the (axial) vector and skew tensor operators of  
twist $\tau$, $O^{(\tau)}_{(5)\alpha}(\kappa_1\lcx, \kappa_2 \lcx)$ and
$M^{(\tau)}_{(5)[\alpha\beta]}(\kappa_1\lcx, \kappa_2 \lcx)$,
are obtained from the (undecomposed)
operators $O_{(5)\alpha}(\kappa_1\lcx, \kappa_2 \lcx)$ and 
 $M_{(5)[\alpha\beta]}(\kappa_1\lcx, \kappa_2 \lcx)$, 
Eqs.~(\ref{O_ent}) and (\ref{M_ent}),
by the application of the corresponding (complete set of) orthogonal 
projectors ({\it including} the $\lambda$--integrations),
${\cal P}^{(\tau)\mu}_\alpha$ and
${\cal P}^{(\tau)[\mu\nu]}_{[\alpha\beta]}$, 
defined by Eqs.~(\ref{O2vec}) -- (\ref{O4vec})
and (\ref{M_tw2_ten}) -- (\ref{M_tw4_ten}), respectively:
\begin{align}
\label{OPROJ}
O^{(\tau)}_{(5)\alpha}(\kappa_1\lcx, \kappa_2 \lcx) &= 
({\cal P}^{(\tau)\mu}_\alpha O_{(5)\mu})(\kappa_1\lcx, \kappa_2 \lcx)
\\
\label{MPROJ}
M^{(\tau)}_{(5)[\alpha\beta]}(\kappa_1\lcx, \kappa_2 \lcx) &= 
({\cal P}^{(\tau)[\mu\nu]}_{[\alpha\beta]} 
M_{(5)[\mu\nu]})(\kappa_1\lcx, \kappa_2 \lcx)
\\
\label{MVPROJ}
M^{(\tau)}_{(5)\alpha}(\kappa_1\lcx, \kappa_2 \lcx) &= 
({\cal P}^{(\tau)\mu}_{(v)\alpha} 
M_{(5)\mu})(\kappa_1\lcx, \kappa_2 \lcx),
\\
\intertext{thereby observing}
\lcx^\alpha M_{(5)\alpha}(\kappa_1\lcx, \kappa_2 \lcx)&=0,
\nonumber
\\
\intertext{with}
\label{OProj}
({\cal P}^{(\tau)}\times{\cal P}^{(\tau')})^{\mu}_\alpha
&=
\delta^{\tau \tau'}
{\cal P}^{(\tau)\mu}_\alpha 
\\
\label{MProj}
({\cal P}^{(\tau)}\times{\cal P}^{(\tau')})^{[\mu\nu]}_{[\alpha\beta]}
&=
\delta^{\tau \tau'}
{\cal P}^{(\tau)[\mu\nu]}_{[\alpha\beta]} .
\end{align}
These projection properties are due to the fact that the Young
operators for different symmetry classes are orthogonal
projections on the tensor space. For hints concerning the proof we refer  
to the Appendix.

In addition, we point to the fact that the twist--2 vector and skew tensor 
operators, Eqs.~(\ref{O2vec}) and (\ref{M_tw2_ten}), are related to the
corresponding scalar and vector operators, (\ref{O2sca}) and (\ref{M_tw2_vec}),
respectively. This leads to relations of the corresponding
distribution functions, for example, cf.~Eqs.~(\ref{matrix_O_t2_sca}) and 
(\ref{matrix_O_t2_nl}) below. 

For convenience, we give the corresponding local expressions of the 
nonlocal operators, Eqs. (\ref{O2sca}) -- (\ref{M_tw3_vec}). 
The relation between the local and nonlocal operators are ($\kappa_1=0$)
\begin{align}
O_\alpha(0, \kappa_2\lcx)
=\sum_{n=0}^\infty \frac{\kappa_2^n}{n!}O_{\alpha n}(\lcx),
\qquad\qquad
M_{[\alpha\beta]}(0, \kappa_2\lcx)
=\sum_{n=0}^\infty \frac{\kappa_2^n}{n!}M_{[\alpha\beta] n}(\lcx).
\nonumber
\end{align}
The expressions for the local operators are 
\begin{align}
\label{O2sca_l}
O^{\rm tw2}_{n+1}(\lcx)
&\equiv
\lcx^\mu O_{\mu n}(\lcx)
=
\bar{\psi}(0)\gamma^5(\gamma\lcx)
(\lcx D)^n \psi(0)
\\
\label{O2vec_l}
O^{\mathrm{tw2}}_{\alpha n}(\lcx)
&=
\hbox{\Large$\frac{1}{n+1}$}
\Big(\pd_\alpha -
\hbox{\Large$\frac{1}{2(n+1)}$}\,x_\alpha\square\Big)
O_{n+1}(x)
\big|_{x=\tilde{x}}
\\
\label{O3vec_l}
O^{\mathrm{tw3}}_{\alpha n}
(\lcx)
&=
\hbox{\Large$\frac{1}{n+1}$}
\Big(n\delta_\alpha^\mu-x^\mu\pd_\alpha
-\hbox{\Large$\frac{1}{n+1}$}
x_\alpha\big((n-1)\pd^\mu
-x^\mu\square\big)
\Big) O_{\mu n}(x)\big|_{x=\tilde{x}}
\\
\label{O4vec_l}
O^{\mathrm{tw4}}_{\alpha n}
(\lcx)
&=\hbox{\Large$\frac{1}{(n+1)^2}$}\,
\lcx_\alpha\Big(n\pd^\mu-\hbox{\Large$\frac{1}{2}$}
x^\mu \square
\Big)
O_{\mu n}( x)\big|_{x=\tilde{x}}\\
\label{M_tw2_ten_l}
M^{\mathrm{tw2}}_{[\alpha\beta]n}(\lcx)
&=\hbox{\Large$\frac{1}{n+2}$}
 \Big\{2\pd_{[\beta}\delta_{\alpha]}^\mu
-\hbox{\Large$\frac{1}{n+1}$}\big(2x_{[\alpha}\pd_{\beta]}\pd^\mu
-x_{[\alpha}\delta_{\beta]}^\mu\square\big)\Big\}
M_{\mu n+1}( x)\big|_{x=\tilde{x}}
\\
\label{M_tw3_ten_l}
M^{\rm tw3}_{[\alpha\beta]n}(\lcx)
&=
\hbox{\Large$\frac{1}{n+2}$}
\Big\{\big(n\delta_{[\beta}^\nu
- 2x^\nu\pd_{[\beta}\big)\delta_{\alpha]}^\mu
+ \hbox{\Large$\frac{2}{n}$}
\Big(
x_{[\alpha}\big((n-1)\delta_{\beta]}^{[\mu} 
-x^{[\mu}\pd_{\beta]}\big)\pd^{\nu]}
-x_{[\alpha}\delta_{\beta]}^{[\mu}x^{\nu]}\square
\nonumber
\\
&\qquad
-x_{[\alpha}\pd_{\beta]}x^{[\mu}\pd^{\nu]}\Big)\Big\}
M_{[\mu\nu] n}(x)\big|_{x=\lcx}
\\
\label{M_tw4_ten_l}
M^{\rm tw4}_{[\alpha\beta]n}(\lcx)
&=
\hbox{\Large$\frac{1}{(n+1)n}$}
\Big\{
x_{[\alpha}\delta_{\beta]}^{[\mu}
x^{\nu]}\square
-2x_{[\alpha}\big((n-1)\delta_{\beta]}^{[\mu} 
-x^{[\mu}\pd_{\beta]}
\big)\pd^{\nu]}\Big\}
M_{[\mu\nu] n}( x)\big|_{x=\lcx}
\\
\label{M_tw2_vec_l}
M^{\mathrm{tw2}}_{\alpha n+1}(\lcx)
&=
M_{\alpha n+1}(\lcx)
-\hbox{\Large$\frac{1}{n+2}$}\,
\lcx_\alpha \pd^\mu
M_{\mu n+1}(x)\big|_{x=\tilde{x}}
\\
\label{M_tw3_vec_l}
M^{\mathrm{tw3}}_{\alpha n+1}(\lcx)
&=
\hbox{\Large$\frac{1}{n+2}$}\,
\lcx_\alpha\pd^\mu
M_{\mu n+1}( x)\big|_{x=\tilde{x}}.
\end{align}

Obviously, analogous projection properties as mentioned above for the 
nonlocal operators also hold for the local ones.
Let us further remark that the projection properties of the above nonlocal 
and local LC operators as well as the conditions of their tracelessness 
can be formulated 
by means of {\em inner derivatives} on the light--cone (see Appendix 
\ref{inner}).

\section{Forward matrix elements of LC--operators with twist $\tau$}
\setcounter{equation}{0}

Now, we define the (polarized) quark distribution functions 
for the bilinear quark operators with definite twist. As usual,
the matrix elements of the nucleon targets, 
$\bar{u}(P,S)\gamma^\mu u(P,S)=2P^\mu$ and 
$\bar{u}(P,S)\gamma^5\gamma^\mu u(P,S)=2S^\mu$, are related to the 
nucleon momentum $P_\mu$ and nucleon spin vector $S_\mu$, respectively,
with $P^2=M^2$, $S^2=-M^2$, $P\cdot S=0$, $M$ denoting the nucleon mass.
Here $u(P,S)$ denotes the free hadronic spinor.

Taking forward matrix elements of Eqs.~(\ref{O2sca}) -- (\ref{M_tw3_vec}) 
we see that, observing the correct tensor structure by the use 
of $P_\mu, S_\mu$ and $\lcx_\mu$, we may introduce any parametrization 
for the matrix elements of the undecomposed operators, 
e.g.,
\begin{align}
\label{FULL}
\langle PS|O_{(5)\alpha}(\kappa_1\lcx,\kappa_2\lcx)|PS\rangle
=2\int_0^1\d z \Big( S_\alpha \widetilde G_1 (z,\mu^2) 
+P_\alpha\frac{\lcx S}{\lcx P} \widetilde G_2(z,\mu^2)
+\lcx_\alpha \frac{\lcx S}{(\lcx P)^2} M^2 \widetilde G_3(z,\mu^2) \Big)
{\e^{\ii\kappa z(\lcx P)}},
\end{align}
where, $\mu^2$ denotes the renormalization scale and, also 
in the following, we defined $\kappa=\kappa_1-\kappa_2$.
Such a choice is not preferable because the matrix element of any
twist--$\tau$ operator then depends in a complicated way from 
every distribution function.
However, according to the above projection properties we are able 
to introduce one (and only one) distribution function for any operator of 
definite twist. 
For cases where the local twist--$\tau$ operators are already known 
in the literature that choice leads to the same result.

Let us first consider the chiral-even {\em pseudo scalar operator} in the 
polarized case. The (forward) matrix element of this twist-2 operator, 
Eq.~(\ref{O2sca}), taken between hadron states $|PS\rangle$ is 
trivially represented as 
\begin{align}
\label{matrix_O_t2_sca}
\langle PS|O_5^{\text{tw2}}(\kappa_1\lcx,\kappa_2\lcx)|PS\rangle
=
2(\lcx S)\int_0^1\d z\, G^{(2)}(z,\mu^2)\,{\e^{\ii\kappa z(\lcx P)}}
=2(\lcx S)\sum_{n=0}^\infty \frac{(\ii\kappa(\lcx P))^n}{n!}G^{(2)}_n(\mu^2) .
\end{align}
Here, $G^{(2)}(z,\mu^2)$ is the uniquely defined twist-2 parton 
distribution function which by a Mellin 
transformation is obtained from the corresponding moments 
\begin{align}
G^{(2)}_n(\mu^2)=\int_0^1\d z\, z^n G^{(2)}(z,\mu^2).
\nonumber
\end{align} 
The {\em nonforward} matrix elements of this scalar operator, 
Eq.~(\ref{O2sca}), are discussed in~\cite{rady97,bgr99,br00}.

Now we consider the chiral-even {\em  axial vector operator}.
Because of the projection properties (\ref{OPROJ}), together
with (\ref{OProj}), we introduce the  parton distribution functions 
$G^{(\tau)}(z, \mu^2)$ of twist $\tau$ by
\begin{align}
\label{Gfct}
\langle PS|O^{(\tau)}_{5\alpha}(\kappa_1\lcx, \kappa_2 \lcx)|PS\rangle 
&\equiv 
\langle PS|({\cal P}^{(\tau)\beta}_\alpha 
O^{(\tau)}_{5\beta})(\kappa_1\lcx, \kappa_2 \lcx)|PS\rangle 
= {\cal P}^{(\tau)\beta}_\alpha
\Big(2 S_\beta\int_0^1\d z\, G^{(\tau)}(z,\mu^2)\,
{\e^{\ii\kappa z(\lcx P)}}\Big)\ .
\end{align} 
For $\tau =2$ this is  consistent with (\ref{matrix_O_t2_sca}), it reads
\begin{align}
\langle PS|O^{\text{tw2}}_{5\alpha}(\kappa_1\lcx,\kappa_2\lcx)|PS\rangle
&=2\int_0^1\d\lambda\Big[\pd_\alpha+\frac{1}{2} \ln(\lambda) 
x_\alpha\square\Big]
(xS)\int_0^1\d z\, G^{(2)}(z,\mu^2)\,{\e^{\ii\kappa \lambda z(xP)}}
\big|_{x=\tilde{x}}\ .
\nonumber
\end{align}
Using the projection operators as they are determined by Eqs.~(\ref{O2vec})
-- (\ref{O4vec}) we obtain (from now on we suppress $\mu^2$)
\begin{align}
\label{matrix_O_t2_nl}
\langle PS|O^{\text{tw2}}_{5\alpha}(\kappa_1\lcx,\kappa_2\lcx)|PS\rangle
&=2\int_0^1\d\lambda\int_0^1\d z\,G^{(2)}(z)
\Big[S_\alpha+\ii\kappa\lambda z 
P_\alpha(\tilde{x}S)
+\frac{\tilde{x}_\alpha}{2}(\tilde{x}S)M^2
(\ii\kappa\lambda z)^2\big(\ln\lambda\big)
\Big]
\e^{\ii\kappa \lambda z(\tilde{x}P)}\\
&=2\sum_{n=0}^\infty \frac{(\ii\kappa(\lcx P))^n}{n!}G^{(2)}_n\Big\{
\frac{1}{n+1}\Big( S_\alpha+n P_\alpha \frac{\lcx S}{\lcx P}\Big)
-\frac{n(n-1)}{2(n+1)^2}\lcx_\alpha\frac{\lcx S}{(\lcx P)^2}M^2\Big\},
\\
\label{matrix_O_t3_nl}
\langle PS|O^{\mathrm{tw3}}_{5\alpha}(\kappa_1\lcx,\kappa_2\lcx)|PS\rangle
&=2\int_0^1\!\!\d\lambda\int_0^1\!\!\d z\, G^{(3)}(z)
\Big[\Big(S_\alpha (\tilde{x}P)
-P_\alpha (\tilde{x}S)\Big)\ii\kappa\lambda z\,
-\tilde{x}_\alpha M^2(\tilde{x}S)
(\ii\kappa\lambda z)^2\ln\lambda\Big]
\e^{\ii\kappa\lambda z(\tilde{x}P)}\\
&=2\sum_{n=1}^\infty \frac{(\ii\kappa(\lcx P))^n}{n!}G^{(3)}_n\Big\{
\frac{n}{n+1}\Big( S_\alpha-P_\alpha \frac{\lcx S}{\lcx P}\Big)
+\frac{n(n-1)}{(n+1)^2}\lcx_\alpha\frac{\lcx S}{(\lcx P)^2}M^2\Big\},
\\
\label{matrix_O_t4_nl}
\langle PS|O^{\mathrm{tw4}}_{5\alpha}(\kappa_1\lcx,\kappa_2\lcx)|PS\rangle
&=
\int_0^1\d\lambda\int_0^1\d z\, G^{(4)}(z)
\Big[\lcx_\alpha(\tilde{x}S)M^2
(\ii\kappa\lambda z)^2\ln\lambda\Big]
\e^{\ii\kappa\lambda z(\tilde{x}P)}\\
&=-\sum_{n=2}^\infty \frac{(\ii\kappa(\lcx P))^n}{n!}G^{(4)}_n
\frac{n(n-1)}{(n+1)^2}\lcx_\alpha\frac{\lcx S}{(\lcx P)^2}M^2.
\end{align}

In the first line of any equation we have written the nonlocal LC-operators
of definite twist, 
and in the second line, after expanding the exponential, we introduced the 
moments of the structure functions; thereby, the $\lambda$--integrations 
contribute the additional $n$-dependent factors. The {\em local} 
twist-2 and twist-3 matrix elements of traceless operators are given 
off--cone in Refs.~\cite{ehrnsperger94,maul97} for $n\leq 3$ and in
Refs.~\cite{picci,blum99} for any $n$. Obviously,
the trace terms which have been explicitly subtracted  
are proportional to $M^2$. According to the terminology of
Jaffe and Ji, they contribute to {\it dynamical} twist-4. 
For the twist--2 operator we observe that the terms proportional to 
$S_\alpha$, $P_\alpha$ and $\lcx_\alpha$ have contributions starting 
with the zeroth, first and second moment, respectively.
The twist--3 operator starts with the first moment, and the
twist--4 operator starts with the second moment. Analogous statements
also hold for the twist--$\tau$ operators below.

Putting together the different twist contributions we obtain,
after replacing $\ii\kappa\lambda z(\lcx P)$ by $\lambda\pd/\pd\lambda$
and performing partial integrations, the following forward
matrix element of the original operator ($\zeta=\kappa(\lcx P)$)
\begin{align}
\label{O_full}
\langle PS|O_{5\alpha}(\kappa_1\lcx,\kappa_2\lcx)|PS\rangle
&=2P_\alpha\frac{\lcx S}{\lcx P}\int_0^1\d z
\Big(G^{(2)}(z)-G^{(3)}(z)\Big)[e_0(\ii\zeta z)-e_1(\ii\zeta z)]\\
&+2S_\alpha\int_0^1\d z
\Big(G^{(2)}(z)e_1(\ii\zeta z)+G^{(3)}(z)
[e_0(\ii\zeta z)-e_1(\ii\zeta z)]\Big)\nonumber\\
&-\lcx_\alpha
\frac{\lcx S}{(\lcx P)^2} M^2\int_0^1\!\!\d z
\Big(G^{(2)}(z)-2G^{(3)}(z)+G^{(4)}(z)\Big)
\Big[e_0(\ii\zeta z)-3e_1(\ii\zeta z)
+2\int_0^1\!\!\d\lambda e_1(\ii\zeta\lambda z)\Big], \nonumber
\end{align}
where we introduced the following ``truncated exponentials''
\begin{align}
e_0(\ii\zeta z)=\e^{\ii\zeta z},\qquad 
e_1(\ii\zeta z)=\int_0^1\!\!\d\lambda\,\e^{\ii\zeta z\lambda}
=\frac{\e^{\ii\zeta z}-1}{\ii\zeta z},
\quad\cdots\quad,
e_{n+1}(\ii\zeta z)
=\frac{(-1)^{n}}{n!}\int_0^1\!\!\d\lambda\,\lambda^n\,\e^{\ii\zeta z\lambda}.
\end{align}
As it should be the application of the projection operators
${\cal P}^{(\tau)\beta}_\alpha$ onto (\ref{O_full}) reproduces
the matrix elements (\ref{matrix_O_t2_nl}) --
(\ref{matrix_O_t4_nl}). In comparison with (\ref{FULL})
we also observe that the distribution functions are accompanied
not simply by the exponentials, $e_0(\ii\zeta z)$, but by more 
involved combinations whose series expansion directly leads to the
representations with the help of moments.

Now we consider the chiral-even {\em vector operator}
$ O_{\alpha}(\kappa_1\lcx,\kappa_2\lcx)=
\bar{\psi}(\kappa_1\lcx)\gamma_{\alpha}
U(\kappa_1\lcx,\kappa_2\lcx)\psi(\kappa_2\lcx)$, which
obeys relations Eqs.~(\ref{O2sca}) -- (\ref{O4vec}), as well as 
(\ref{OPROJ}) and (\ref{OProj}) with the {\em same} projection operator
as the axial vector operator. Let us introduce the corresponding
parton distribution functions $F^{(\tau)}(z)$ of twist $\tau$ by
\begin{align}
\label{Ffct}
\langle PS| O^{(\tau)}_\alpha(\kappa_1\lcx, 
\kappa_2 \lcx)|PS\rangle 
&= {\cal P}^{(\tau)\mu}_\alpha\Big(
2 P_\mu\int_0^1\d z\, F^{(\tau)}(z)\,{\e^{\ii\kappa z(xP)}}
\Big).
\end{align}
Again, this is consistent with the definition of the unpolarized 
distribution function $F^{(2)}(z)$ by the matrix element of the 
corresponding twist-2 {\em scalar operator} 
\begin{align}
\label{matrix_O_tw2_sca}
\langle P| O^{\text{tw2}}(\kappa_1\lcx,\kappa_2\lcx)|P\rangle
=
2(\lcx P)\int_0^1\d z\, F^{(2)}(z)\,\e^{\ii\kappa z(\lcx P)}
=2(\lcx P)\sum_{n=0}^\infty \frac{(\ii\kappa(\lcx P))^n}{n!}F^{(2)}_n .
\end{align}
The {\em nonforward} matrix elements of this scalar operator 
has been considered for the first time in~\cite{mul94}.

The forward matrix elements of the {\em vector operators} of twist $\tau$
are obtained as follows:
\begin{align}
\label{matrix_O_tw2_nl}
\hspace{-.3cm}
\langle P|
O^{\text{tw2}}_{\alpha}(\kappa_1\lcx,\kappa_2\lcx)|P\rangle
&=2\int_0^1\!\!\d\lambda\!\!\int_0^1\!\!\d z\,F^{(2)}(z)
\Big[P_\alpha\big(1+\ii\kappa\lambda z(\tilde{x}P)\big)
+\lcx_\alpha M^2\ii\kappa\lambda z\ln\lambda
\Big(1+\frac{1}{2}\ii\kappa\lambda z(\tilde{x}P)\Big)\Big]
\e^{\ii\kappa \lambda z(\tilde{x}P)}\\
\hspace{-.3cm}
&=2\sum_{n=0}^\infty \frac{(\ii\kappa(\lcx P))^n}{n!}F^{(2)}_n\Big\{
P_\alpha -\frac{n}{2(n+1)}\lcx_\alpha\frac{M^2}{\lcx P}\Big\},
\\
\hspace{-.3cm}
\label{matrix_O_tw3_nl}
\langle P|
 O^{\mathrm{tw3}}_{\alpha}(\kappa_1\lcx,\kappa_2\lcx)|P\rangle
&=-2\lcx_\alpha M^2\int_0^1\!\!\d\lambda
\int_0^1\!\!\d z\, F^{(3)}(z)
\ii\kappa\lambda z\,\Big[
\Big(1+2\ln\lambda\big)
+\ii\kappa\lambda z\big(\ln\lambda\big)
\left(\tilde{x}P\right)\Big]
\e^{\ii\kappa\lambda z(\tilde{x}P)} 
\equiv 0\,,
\\
\hspace{-.3cm}
\langle P|
O^{\mathrm{tw4}}_{\alpha}(\kappa_1\lcx,\kappa_2\lcx)|P\rangle
&=-2\lcx_\alpha M^2
\int_0^1\!\!\d\lambda\int_0^1\!\!\d z\,F^{(4)}(z)
\ii\kappa\lambda z\, \big(\ln\lambda\big)
\Big(1+\frac{1}{2}\ii\kappa\lambda z(\tilde{x}P)\Big)
\e^{\ii\kappa \lambda z(\tilde{x}P)}
\\
\hspace{-.3cm}
&=\sum_{n=1}^\infty \frac{(\ii\kappa(\lcx P))^n}{n!}F^{(4)}_n
\frac{n}{(n+1)}\lcx_\alpha\frac{M^2}{\lcx P},
\end{align}

The off--cone traceless local twist-2 matrix element has already been 
obtained in~\cite{guth,georgi}.
The vanishing of the twist-3 structure function $F^{(3)}(z)$ which
follows here by partial integration proves in the {\em nonlocal} case
the same fact that Jaffe and Soldate have already proved for the 
{\em local} twist-3 operator~\cite{JS82}. 
The trace terms of the twist-2 operator and the twist-4 
operator itself both contribute to dynamical twist-4 and, consequently, 
to the same $1/Q^2$-behaviour in the cross section. 

Putting together all the contributions of different twist and 
performing partial $\lambda$--integrations, we obtain the following 
representation for the forward matrix element of the original operator
\begin{align}
\label{matrix_O_nl}
&\langle P| O^{}_{\alpha}(\kappa_1\lcx,\kappa_2\lcx)|P\rangle
=2P_\alpha\int_0^1\d z\, F^{(2)}(z) e_0(\ii\zeta z)
-
\lcx_\alpha\frac{M^2}{\lcx P}\int_0^1\d z
\Big(F^{(2)}(z)-F^{(4)}(z)\Big)
\big[e_0(\ii\zeta z)- e_1(\ii\zeta z)\big].
\end{align}

The matrix element of the  simplest bilocal {\em scalar operator} 
arises as
\begin{align}\label{}
\langle P|\bar{\psi}(\kappa_1\lcx)
U(\kappa_1\lcx,\kappa_2\lcx)\psi(\kappa_2\lcx)|P\rangle
=2M\int_0^1\d z\, E^{(3)}(z){\e^{\ii\kappa z(\lcx P)}}
=2M\sum_{n=0}^\infty \frac{(\ii\kappa(\lcx P))^n}{n!}E^{(3)}_n ,
\end{align}
where $E^{(3)}(z)$ is another spin-independent twist-3 structure 
function.

Now, we consider the matrix elements of the chiral-odd {\em  vector 
operators}. In fact they may be obtained from the chiral-odd skew-tensor 
operator by contraction with $\lcx_\beta$ (cf. Eq.~(\ref{M_vec})). 
This also determines the 
corresponding structure functions which are introduced by
\begin{align}
\label{Hfct}
\langle PS|M^{(\tau)}_{5[\alpha\beta]}(\kappa_1\lcx, \kappa_2 \lcx)|PS\rangle 
&=\frac{2}{M}{\cal P}^{(\tau)[\mu\nu]}_{[\alpha\beta]}
\Big( (S_\mu P_\nu-S_\nu P_\mu) 
\int_0^1\d z\, H^{(\tau)}(z)\,{\e^{\ii\kappa  z(\lcx P)}}
\Big).
\end{align}
Applying the same procedure as above we obtain
\begin{align}
\label{matrix_Mv_tw2}
\langle PS|M^{\mathrm{tw2}}_{5\alpha}(\kappa_1\lcx,\kappa_2\lcx)|PS\rangle
&=
\frac{2}{M}\int_0^1\!\!\d z H^{(2)}(z)\!
\left[\Big(S_{\alpha}(\tilde{x}P)-P_\alpha(\tilde{x}S)\Big)
\e^{\ii\kappa  z(\tilde{x}P)}
+\tilde{x}_\alpha(\tilde{x}S)M^2
\!\!\int_0^1\!\! \d \lambda\,\ii\kappa z\lambda^2 \, 
\e^{\ii\kappa \lambda z(\tilde{x}P)}\right],\\
&=\frac{2}{M}\sum_{n=0}^\infty \frac{(\ii\kappa(\lcx P))^n}{n!}H^{(2)}_n\Big\{
 S_\alpha(\lcx P)- P_\alpha (\lcx S)
+\frac{n}{n+2}\lcx_\alpha\frac{\lcx S}{\lcx P} M^2\Big\},
\\
\label{matrix_Mv_tw3}
\langle PS|M^{\mathrm{tw3}}_{5\alpha}(\kappa_1\lcx,\kappa_2\lcx)|PS\rangle
&=-2\int_0^1\!\!\d\lambda
\int_0^1\!\! \d z\,H^{(3)}(z)
\Big[\tilde{x}_\alpha (\tilde{x}S)M\ii\kappa z \lambda^2\Big]
\e^{\ii\kappa \lambda z(\tilde{x}P)}\\
&=-2\sum_{n=1}^\infty \frac{(\ii\kappa(\lcx P))^n}{n!}H^{(3)}_n
\frac{n}{n+2}\lcx_\alpha\frac{\lcx S}{\lcx P} M.
\end{align}
The twist--2 part is in agreement with Jaffe and Ji's local 
expression~(see Eq.~(45) in \cite{Jaf92}) and also with~\cite{Koike95}. 
Adding up both local terms the matrix element of 
$M_{\alpha}(\kappa_1\lcx,\kappa_2\lcx)$ 
coincides with the corresponding one of \cite{Jaf92}; its nonlocal 
version reads:
\begin{align}\label{M_vector_matrix}
\langle PS|M_{5\alpha}(\kappa_1\lcx,\kappa_2\lcx)|PS\rangle
=&\,
\frac{2}{M}\Big(S_{\alpha}(\tilde{x}P)-P_\alpha(\tilde{x}S)\Big)
\int_0^1\!\!\d z H^{(2)}(z)\, e_0(\ii\zeta z)
\nonumber\\
&+\tilde{x}_\alpha \frac{\tilde{x}S}{\lcx P} M
\int_0^1\!\!\d z \Big(H^{(2)}(z)-H^{(3)}(z)\Big)
\Big( e_0(\ii\zeta z) +2 e_2(\ii\zeta z)\Big).
\end{align}
Obviously, the twist-2 distribution function $H^{(2)}(z)$ and 
the twist-3 distribution function $H^{(3)}(z)$ contribute 
destructive to the same $1/Q^2$-behaviour of the cross section.
A similar {\em nonlocal} twist--2 matrix element (modulo correction
of the trace term) has been given in~\cite{Kodaira99}.

Now, we consider the more involved chiral-odd {\em  skew tensor operator}. 
The matrix elements of the skew tensor operators of twist $\tau$ 
are obtained using the projectors determined by Eqs.~(\ref{M_tw2_ten})
-- (\ref{M_tw4_ten}):
\begin{align}
\label{matrix_M_tw2_nl}
&\langle PS|
M^{\mathrm{tw2}}_{5[\alpha\beta]}(\kappa_1\lcx,\kappa_2\lcx)|PS\rangle
=\frac{2}{M}\int_0^1\d\lambda\int_0^1\d z\, H^{(2)}(z)
\Big[2\lambda\, S_{[\alpha} P_{\beta]}\big(2+\ii\kappa\lambda z(\lcx P)\big)
\nonumber\\
&\qquad\qquad\qquad\qquad\qquad\qquad\qquad
+\left(1-\lambda\right)M^2
\lcx_{[\alpha}\Big\{4(\ii\kappa\lambda z)S_{\beta]}
+(\ii\kappa\lambda z)^2\big(S_{\beta]}(\lcx P)
+P_{\beta]}(\lcx S)\big)\Big\}
\Big]\e^{\ii\kappa\lambda z(\lcx P)}
\\
&\quad
=\frac{2}{M}\sum_{n=0}^\infty \frac{(\ii\kappa(\lcx P))^n}{n!}H^{(2)}_n\Big\{
 2S_{[\alpha}P_{\beta]}
+\frac{4n}{(n+2)(n+1)}\frac{M^2}{\lcx P}\lcx_{[\alpha}S_{\beta]}
+\frac{n(n-1)}{(n+2)(n+1)}\frac{M^2}{\lcx P}\lcx_{[\alpha}\Big(S_{\beta]}
+P_{\beta]}\frac{\lcx S}{\lcx P}\Big)\Big\},
\\
&\langle PS|
M^{\mathrm{tw3}}_{5[\alpha\beta]}(\kappa_1\lcx,\kappa_2\lcx)|PS\rangle
=2\int_0^1\d\lambda\frac{1-\lambda^2}{\lambda}
\int_0^1\d z\,\ii\kappa\lambda z H^{(3)}(z) M
\lcx_{[\alpha}\Big\{ S_{\beta]}
+\ii\kappa\lambda  z P_{\beta]}(\lcx S)\Big\}
\e^{\ii\kappa\lambda z(\lcx P)}\\
&\phantom{\langle PS|
M^{\mathrm{tw2}}_{\alpha\beta}(\kappa_1\lcx,\kappa_2\lcx)|PS\rangle}
=-2\sum_{n=1}^\infty \frac{(\ii\kappa(\lcx P))^n}{n!}H^{(3)}_n
\frac{2}{n+2}\frac{M}{\lcx P}\lcx_{[\alpha}\Big(S_{\beta]}
+(n-1)P_{\beta]}\frac{\lcx S}{\lcx P}\Big),
\\
\label{matrix_M_tw4_nl}
&\langle PS|
M^{\mathrm{tw4}}_{5[\alpha\beta]}(\kappa_1\lcx,\kappa_2\lcx)|PS\rangle
=-2\int_0^1\d\lambda\,\frac{1-\lambda}{\lambda}
\int_0^1\d z\, (\ii \lambda\kappa z)^2 H^{(4)}(z)M
\lcx_{[\alpha}\Big\{S_{\beta]}(\lcx P)-P_{\beta]}(\lcx S)\Big\}
\e^{\ii\kappa\lambda z(\lcx P)}\\
&\phantom{\langle PS|
M^{\mathrm{tw2}}_{[\alpha\beta]}(\kappa_1\lcx,\kappa_2\lcx)|PS\rangle}
=-2\sum_{n=2}^\infty \frac{(\ii\kappa(\lcx P))^n}{n!}H^{(4)}_n
\frac{n-1}{n+1}\frac{M}{\lcx P}\lcx_{[\alpha}\Big(S_{\beta]}
-P_{\beta]}\frac{\lcx S}{\lcx P}\Big).
\end{align}

Again, the moments of distributions functions of twist $\tau = 2,3$
and $4$ begin with $n=0,1$ and $2$, respectively. In addition,
we remark that only those terms of the operator (\ref{M_tw3_ten}) 
contribute to the twist--3 structure function which result from
the trace terms of (\ref{M_tw2_ten}). Analogous to the vector case
the forward matrix element of the `true' twist--3 part of (\ref{M_tw3_ten})
vanishes. 
In Eq.~(\ref{matrix_M_tw4_nl}) 
only the twist--4 operator contributes which result from
the trace terms of (\ref{M_tw2_ten}).
Let us also mention that after multiplication of (\ref{matrix_M_tw4_nl}) with
$\lcx_\alpha$ (or $\lcx_\beta$) the matrix element vanishes because the 
corresponding vector operator does not contain any twist-four contribution.
This is a simple but important property which may be traced back to the fact
that the corresponding Young pattern $[n+2] = (n,1,1)$ does allow only 
$n$ symmetrizations; it is therefore characteristic for any twist--4 
skew tensor operator.  

The matrix element of the original skew tensor operator is obtained as
\begin{align}
\label{M_full_JJ}
\langle PS|M_{5[\alpha\beta]}(\kappa_1\lcx,\kappa_2\lcx)|PS\rangle
&=
\frac{4}{M}S_{[\alpha}P_{\beta]}
\int_0^1\d z H^{(2)}(z)e_0(\ii\zeta z)\\
&+2\lcx_{[\alpha}P_{\beta]}\frac{\lcx S}{(\lcx P)^2} M
\int_0^1\d z \Big\{
  H^{(2)}(z)\big[e_0(\ii\zeta z) + 2e_1(\ii\zeta z) + 6e_2(\ii\zeta z)\big]
\nonumber\\
&\qquad\qquad\qquad\qquad
- H^{(3)}(z)\big[1 +2e_0(\ii\zeta z) + 6e_2(\ii\zeta z)\big]
+ H^{(4)}(z)\big[1 + e_0(\ii\zeta z) - 2e_1(\ii\zeta z)\big]
\Big\}\nonumber\\
&
+2\lcx_{[\alpha}S_{\beta]}\frac{M}{\lcx P}
\int_0^1\d z \Big\{
  H^{(2)}(z)\big[e_0(\ii\zeta z)-2e_1(\ii\zeta z)-2e_2(\ii\zeta z)\big]
\nonumber\\
&\qquad\qquad\qquad\qquad
+ H^{(3)}(z)\big[1+2e_2(\ii\zeta z)\big]
- H^{(4)}(z)\big[1+e_0(\ii\zeta z)-2e_1(\ii\zeta z)\big]
\Big\}.
\nonumber
\end{align}

This finishes the determination of the nine nontrivial distribution
functions which result from the forward matrix elements of the nonlocal
light--cone quark operators of definite twist.

\section{Relations between new and conventional distribution functions}
\setcounter{equation}{0}

Obviously, since these new distribution functions are related to true
traceless operators they differ from the conventional ones \cite{Jaf92}
for twist $\tau \geq 3$ at least by the contributions from the trace terms.
As far as the scalar LC--operators are concerned which definitely are
of twist--2 the new and the old distributions functions coincide.
However, for the vector and (skew) tensor operators also the 
contributions of dynamical twist--2 differ from those of geometric
twist--2.

In order to be able to compare conventional and new structure functions,
we rewrite the matrix elements (\ref{O_full}), (\ref{matrix_O_nl}) and
(\ref{M_full_JJ}) by choosing
\begin{align}
\lcx_\alpha = x_\alpha -\frac{P_\alpha}{M^2} \Big(
(xP) - \sqrt{(xP)^2 - x^2 M^2}
\Big), \qquad
P_\alpha=p_\alpha+\frac{1}{2} \lcx_\alpha\frac{M^2}{\lcx P},\qquad
S_\alpha=S^\bot_\alpha+p_\alpha \frac{\lcx S}{\lcx P}-
\frac{1}{2} \lcx_\alpha \frac{\lcx S}{(\lcx P)^2} M^2,
\end{align}
where $p_\alpha$ is a light-like vector ($p^2=0$ and $p\cdot\lcx=P\cdot\lcx$)
and $S^\bot_\alpha$ is the transversal spin-polarization. 
Using this parametrization we obtain
\begin{align}
\label{JJ_full}
&\langle PS|O_{5\alpha}(\kappa_1\lcx,\kappa_2\lcx)|PS\rangle
=2p_\alpha\frac{\lcx S}{\lcx P}\!\int_0^1\!\!\d z\, G^{(2)}(z)e_0(\ii\zeta z)
+2S^\bot_\alpha\!\!\int_0^1\!\!\d z
\Big\{G^{(2)}(z)e_1(\ii\zeta z)+G^{(3)}(z)
[e_0(\ii\zeta z)-e_1(\ii\zeta z)]\Big\}
\nonumber\\
&\qquad
-\lcx_\alpha\frac{\lcx S}{(\lcx P)^2}M^2\int_0^1\!\!\d z
\Big\{G^{(4)}(z)\Big[e_0(\ii\zeta z)-3e_1(\ii\zeta z)
+2\int_0^1\!\!\d\lambda\, e_1(\ii\zeta\lambda z)\Big]
-G^{(2)}(z)\Big[e_1(\ii\zeta z)
-2\int_0^1\!\!\d\lambda\, e_1(\ii\zeta\lambda z)\Big]
\nonumber\\
&\qquad\qquad\qquad
+4G^{(3)}(z)\Big[e_1(\ii\zeta z)
-\int_0^1\!\!\d\lambda\, e_1(\ii\zeta\lambda z )\Big]\Big\},
\\
\label{JJ_O_nl}
&\langle P| O^{}_{\alpha}(\kappa_1\lcx,\kappa_2\lcx)|P\rangle
=2p_\alpha\int_0^1\!\!\d z\, F^{(2)}(z) e_0(\ii\zeta z)
+\lcx_\alpha\frac{M^2}{\lcx P}\int_0^1\!\!\d z
\Big\{F^{(4)}(z)e_0(\ii\zeta z)+
\big[F^{(2)}(z)-F^{(4)}(z)\big]e_1(\ii\zeta z)\Big\} ,
\\
&\langle P|\bar{\psi}(\kappa_1\lcx)
U(\kappa_1\lcx,\kappa_2\lcx)\psi(\kappa_2\lcx)|P\rangle
=2M\int_0^1\d z\, E^{(3)}(z){e_0{(\ii\zeta z)}},
\\
\label{JJM_full}
&\langle PS|M_{5[\alpha\beta]}(\kappa_1\lcx,\kappa_2\lcx)|PS\rangle
=
\frac{4}{M}S^\bot_{[\alpha}p_{\beta]}
\int_0^1\!\!\d z H^{(2)}(z)e_0(\ii\zeta z)
\\
&\qquad
+2\lcx_{[\alpha}p_{\beta]}\frac{\lcx S}{(\lcx P)^2}M
\int_0^1\!\!\d z \Big\{
 4H^{(2)}(z)e_2(\ii\zeta z)
-2H^{(3)}(z)\big[e_0(\ii\zeta z)+2e_2(\ii\zeta z)\big]
\Big\}
\nonumber\\
&\qquad
-2\lcx_{[\alpha}S^\bot_{\beta]}\frac{M}{\lcx P}
\int_0^1\!\!\d z \Big\{
2 H^{(2)}(z)\big[e_1(\ii\zeta z)+e_2(\ii\zeta z)\big]
- H^{(3)}(z)\big[1+2e_2(\ii\zeta z)\big]
+ H^{(4)}(z)\big[1+e_0(\ii\zeta z)-2e_1(\ii\zeta z)\big]
\Big\} .\nonumber
\end{align}
Let us note that after multiplication of (\ref{JJM_full}) with $\lcx_\alpha$
(or $\lcx_\beta$) the last part of the matrix element vanishes 
($\lcx\cdot S^\bot=0$). This means that the dynamical twist-4 function
$h_3(z)$ can be ignored in the case of the vector operator 
$M_{5\alpha}(\kappa_1\lcx,\kappa_2\lcx)$ but not in the case of the 
more general tensor operator $M_{5[\alpha\beta]}(\kappa_1\lcx,\kappa_2\lcx)$.

On the other hand, these matrix elements
in terms of the nine parton distributions introduced 
by Jaffe and Ji~\cite{Jaf92} are given as follows:
\begin{align}
&\langle PS|O_{5\alpha}(\kappa_1\lcx,\kappa_2\lcx)|PS\rangle  
= 2\left[ p_{\alpha}
\frac{\lcx S}{\lcx P}
\int_{0}^{1} \!\d z\, \e^{\ii\zeta z} g_{1}(z) 
+ S^{\perp}_\alpha\int_{0}^{1} \!\d z\, \e^{\ii\zeta z} g_{T}(z) 
+ \lcx_{\alpha}
\frac{\lcx S}{(\lcx P)^{2}} M^{2}
\int_{0}^{1} \!\d z\,\e^{\ii\zeta z} g_{3}(z) \right],
\label{eq:fvda}
\\
&\langle P|O_{\alpha}(\kappa_1\lcx,\kappa_2\lcx)|P\rangle
 =  2 \left[ p_\alpha \int_{0}^{1} \!\d z\, \e^{\ii\zeta z}
f_1(z) + \lcx_\alpha \frac{M^2}{\lcx P}  \int_{0}^{1} \!\d z\, 
\e^{\ii\zeta z} f_4(z)\right],
\\
&\langle P|\bar{\psi}(\kappa_1\lcx)
U(\kappa_1\lcx,\kappa_2\lcx)\psi(\kappa_2\lcx)|P\rangle
= 2M \int_{0}^{1} \!\d z\, \e^{\ii\zeta z} e(z),\\
&\langle PS|M_{5[\alpha\beta]}(\kappa_1\lcx,\kappa_2\lcx)|PS\rangle 
=\frac{4}{M}\left[S^{\perp}_{[\alpha} p_{\beta]}
\int_{0}^{1} \!\!\d z\, \e^{\ii \zeta z} h_{1}(z) 
-\lcx_{[\alpha} p_{\beta]}
\frac{\lcx S M^2}{(\lcx P)^{2}}\! \!
\int_{0}^{1} \!\!\d z\, \e^{\ii\zeta z} h_{L}(z) 
-\lcx_{[\alpha}S^{\perp}_{\beta]} 
\frac{M^{2}}{\lcx P}  \!\!
\int_{0}^{1} \!\!\d z\, \e^{\ii\zeta z} h_{3}(z) \right].
\label{eq:ftda}
\end{align}

From the expressions (\ref{JJ_full}) -- (\ref{JJM_full}) and 
(\ref{eq:fvda}) -- (\ref{eq:ftda})
it is obvious that the conventional structure            
functions of twist $t \geq 3$ contain contributions also of lower
geometric twist. In Tab.~\ref{tab:1} we classify them according to  
their spin, dynamical twist and chirality and we also note their
content of geometric twist: 
\begin{table}
\begin{center}
\renewcommand{\arraystretch}{1.3}
\begin{tabular}{|c|ccc|}
\hline
Twist $t$   & 2 & 3 & 4 \\
    & $O(1)$  & $O(1/Q)$& $O(1/Q^{2})$ \\ \hline
spin ave.& $f_{1}=F^{(2)}$ & $\underline{e}=E^{(3)}$ & $f_{4}=F^{(4)}/2+F_f(F^{(2)},F^{(4)})/2$ \\
$S_{\parallel}$ & $g_{1}=G^{(2)}$ & $\underline{h_{L}}=H^{(3)}+F_L(H^{(2)},H^{(3)})$ & $g_{3}=-G^{(4)}/2+F_g(G^{(2)},G^{(3)},G^{(4)})/2$ \\
$S_{\perp}$ & $\underline{h_{1}}=H^{(2)}$ & $g_{T}=G^{(3)}+F_T(G^{(2)},G^{(3)})$ & $\underline{h_{3}}=H^{(4)}/2+F_h(H^{(2)},H^{(3)},H^{(4)})/2$
\\[2pt]
\end{tabular}
\renewcommand{\arraystretch}{1}
\end{center}
\caption{Spin, dynamical twist and chiral classification of the 
nucleon structure functions.}
\label{tab:1}
\end{table}%
The parton distributions in 
the first row are spin-independent, those in the
second and third row describe
longitudinally ($S_{\parallel}$) and transversely ($S_{\perp}$)
polarized nucleons, respectively.
The underlined distributions are referred to as 
chiral-odd, 
because they correspond to 
chirality-violating Dirac matrix structures 
$\Gamma = \{\sigma_{\alpha\beta} \ii \gamma_{5},\, 1\}$. 
The other distributions are termed chiral-even, because of the 
chirality-conserving structures 
$\Gamma = \{\gamma_{\alpha},\, \gamma_\alpha\gamma_5\}$. 

Re-expressing the truncated exponentials
and performing appropriate variable transformations
we obtain the following relations, which give the interrelation between the 
structure functions of different twist:
\begin{align}  
\label{rel-g1}
g_1(z)&=G^{(2)}(z),\\
\label{rel-gT}
g_T(z)&=G^{(3)}(z) + \int_z^1 \frac{\d y}{y}
\Big(G^{(2)}-G^{(3)}\Big)\!\left(y\right),\\
\label{rel-g3}
2g_3(z)&=-G^{(4)}(z) + \int_z^1 \frac{\d  y}{y}\Big\{
\Big(G^{(2)}-4G^{(3)}+3G^{(4)}\Big)\!\left(y\right)
+ 2 \ln \Big(\frac{z}{y}\Big)
\Big(G^{(2)}-2G^{(3)}+G^{(4)}\Big)\!\left(y\right)
\Big\},\\
\label{rel-f1}
f_1(z)&=F^{(2)}(z),\\
\label{rel-f4}
2f_4(z)&=F^{(4)}(z) + \int_z^1 \frac{\d y}{y}
\Big(F^{(2)}-F^{(4)}\Big)\!\left(y\right),\\
e(z)&=E^{(3)}(z),\\
\label{rel-h1}
h_1(z)&=H^{(2)}(z),\\
\label{rel-hL}
h_L(z)&=H^{(3)}(z) + 2z \int_z^1 \frac{\d y}{y^2}
\Big(H^{(2)}-H^{(3)}\Big)\!\left(y\right),\\
\label{rel-h3}
2h_3(z)&=H^{(4)}(z)+ \int_z^1 \frac{\d y}{y}\Big\{
2\Big(H^{(2)}-H^{(4)}\Big)\!\left(y\right)
-2\frac{z}{y}\Big(H^{(2)}-H^{(3)}\Big)\left(y\right)
-\delta\Big(\frac{z}{y}\Big)\Big(H^{(3)}-H^{(4)}\Big)\!\left(y\right)
\Big\}.
\end{align}
These relations between the conventional and the new quark distribution
functions hold for $1 \geq z \geq 0$; the corresponding antiquark
distribution functions are obtained for $z \rightarrow -z$. Furthermore, 
we observe that both decompositions coincide in the leading terms, but 
differ at higher order. 
For instance, $g_1(z)$ and $h_1(z)$ are genuine geometric 
twist-2 structure functions and $e(z)$ is a genuine twist-3 function.
Moreover, the parton distribution functions 
$g_T(z)$, $f_1(z)$ and $h_L(z)$ with dynamical twist $t=3$ also contain 
contributions 
of geometrical twist $\tau = 2$ and $3$. Additionally, the dynamical twist
$t=4$ functions $g_3(z)$ and $h_3(z)$ contain geometrical twist-2, twist-3 
as well as twist-4 parts and the dynamical twist
$t=4$ function $f_4(z)$ contains geometrical twist-2 and twist-4.

These relationship may be inverted. The nontrivial inverse relations are:
\begin{align}
\label{rel-G3}
G^{(3)}(z)&=g_T(z) +\frac{1}{z} \int_z^1 \d y
\big(g_T-g_1\big)\!\left(y\right),\\
G^{(4)}(z)&=-\Big\{2g_3(z) 
+\frac{1}{z^2} \int_z^1 \d y\, y
\big(6g_3+4g_T-g_1\big)\!\left(y\right)
-\frac{1}{z^2} \int_z^1 \d y\, y \ln\Big(\frac{z}{y}\Big)
\big(2g_3+4g_T-3g_1\big)\!\left(y\right)\Big\},\\
F^{(4)}(z)&=2f_4(z) + \frac{1}{z}\int_z^1 \d y
\big(2f_4-f_1\big)\!\left(y\right),\\
\label{rel-H3}
H^{(3)}(z)&=h_L(z) + \frac{2}{z}\int_z^1 \d y
\big(h_L-h_1\big)\!\left(y\right)\,\\
\label{rel-H4}
H^{(4)}(z)&=2\Big\{h_3(z) 
+\frac{1}{z^2} \int_z^1 \d y\, y
\big(2h_3-h_L\big)\!\left(y\right)
+\frac{1}{z^2} \int_z^1 \d y\, y \ln\Big(\frac{z}{y}\Big)
\big(h_L-h_1\big)\!\left(y\right)\Big\}.
\end{align}

The relation between the moments may be read off from Eqs.~(\ref{JJ_full})
-- (\ref{JJM_full}) as follows:
\begin{align}  
g_{1n}&=G^{(2)}_n,\\
g_{Tn}&=G^{(3)}_n +  \frac{1}{n+1}
\Big(G^{(2)}_n-G^{(3)}_n\Big),\\
2g_{3n}&=-G^{(4)}_n +  \frac{1}{n+1}
\Big(G^{(2)}_n-4G^{(3)}_n+3G^{(4)}_n\Big)
-\frac{2}{(n+1)^2}
\Big(G^{(2)}_n-2G^{(3)}_n+G^{(4)}_n\Big),\\
f_{1n}&=F^{(2)}_n,\\
2f_{4n}&=F^{(4)}_n +  \frac{1}{n+1}
\Big(F^{(2)}_n-F^{(4)}_n\Big),\\
e_n&=E^{(3)}_n,\\
h_{1n}&=H^{(2)}_n,\\
h_{Ln}&=H^{(3)}_n +  \frac{2}{n+2}
\Big(H^{(2)}_n-H^{(3)}_n\Big),\\
2h_{3n}&=H^{(4)}_n+  \frac{2}{n+1}
\Big(H^{(2)}_n-H^{(4)}_n\Big)
-\frac{2}{n+2}\Big(H^{(2)}_n-H^{(3)}_n\Big)
-\delta_{n0}\Big(H^{(3)}_n-H^{(4)}_n\Big).
\end{align}
In terms of the moments the relations between conventional and new distribution
functions may be easily inverted. The nontrivial inverse relations are:
\begin{align}  
G^{(3)}_{n}&=g_{Tn} +  \frac{1}{n}
\Big(g_{Tn}-g_{1n}\Big),& n>0\\
G^{(4)}_{n}&=-
\Big\{ 2g_{3n}+  \frac{1}{n-1}
\Big(6g_{3n}+4g_{Tn}-g_{1n}\Big)
+\frac{1}{n(n-1)}\Big(2g_{3n}+4g_{Tn}-3g_{1n}\Big)\Big\},& n>1\\
F^{(4)}_n&=2f_{4n} +  \frac{1}{n}
\Big(2f_{4n}-f_{1n}\Big),& n>0\\
H^{(3)}_{n}&=h_{Ln}+  \frac{2}{n}
\Big(h_{Ln}-h_{1n}\Big),& n>0\\
H^{(4)}_n&=
2\Big\{ h_{3n}+  \frac{1}{n-1}
\Big(2h_{3n}-h_{Ln}\Big)
-\frac{1}{n(n-1)}\Big(h_{Ln}-h_{1n}\Big)\Big\},& n>1.
\end{align}

The nontrivial relationships between the conventional and the new
distribution functions are much simpler than to be assumed by a first
glance at the expressions like (\ref{O_full}), (\ref{matrix_O_nl}) and
(\ref{M_full_JJ}). They show that the conventional distribution functions 
are determined by the new ones of the same as well as lower geometrical 
twist, and vice versa (with respect to dynamical twist).
Obviously, the same holds for their moments. In principle, this
allows to determine, e.g., the new distribution amplitudes from
the experimental data if these are known for the conventional
ones. At least, this should be possible for the lowest moments.

\section{Wandzura-Wilczek--like relations}
\setcounter{equation}{0}

As an immediate consequence of the interrelations (\ref{rel-g1}) --
(\ref{rel-h3}) and (\ref{rel-G3}) -- (\ref{rel-H4}) between the conventional 
structure functions and the new ones of genuine geometric twist
we are able to derive the decomposition of the conventional structure
functions into its parts of genuine twist. Thereby, we also obtain new
Wandzura-Wilczek--like relations for the conventional structure functions
and also new sum rules of the type of
the Burkhardt-Cottingham sum rule.

Let us introduce the usual notation $g_T(z)=g_1(z)+g_2(z)$ and the following
decomposition of $g_2(z)=g^{(2)}_2(z)+g^{(3)}_2(z)$ into its parts
$g^{(2)}_2(z)$ and $g^{(3)}_2(z)$ of genuine twist-2 and twist-3,
respectively, part of $g_2(z)$. Then,
substituting (\ref{rel-g1}) into (\ref{rel-gT}), we get
\begin{align}
\label{WW-tw2}
g^{(2)}_2(z)&=-g_1(z) + \int_z^1 \frac{\d y}{y}g_1(y)\\
\label{WW-tw3}
g^{(3)}_2(z)&=g_2(z) + g_1(z)- \int_z^1 \frac{\d y}{y}g_1(y),
\end{align}
where (\ref{WW-tw2}) is just the Wandzura-Wilczek relation 
for the twist-2 part~\cite{Wandzura}. On the other hand, the 
analogous relation (\ref{WW-tw3}) for the twist-3 part, which is an
immediate consequence of the above definition, has already been obtained 
in the framework of the local OPE~\cite{blum97}
and the nonlocal OPE~\cite{Lazar98}.
For the moments we obtain
\begin{align}
\label{WW-tw2-n}
g^{(2)}_{2n}&=-g_{1 n}+\frac{1}{n+1}\,g_{1 n},\\
\label{WW-tw3-n}
g^{(3)}_{2 n}&=g_{2 n} + \frac{n}{n+1}\, g_{1n },\qquad n>0.
\end{align}
Obviously, from the relation (\ref{WW-tw2-n}) for $n=0$ the 
Burkhardt- Cottingham sum rule~\cite{BC70} follows:
\begin{align}
\int_0^1\d z\, g_2(z)=0.
\end{align}

Of course, these relationships are well known. However, with the help of
the results of the preceeding chapter we are able to generalize these
relationships also to the other nontrivial structure functions.
Using the formulas (\ref{rel-g1}), (\ref{rel-gT}) (\ref{rel-g3}) and
(\ref{rel-G3}), we obtain Wandzura-Wilczek--like integral relations for 
geometric twist parts of
the function $g_3(z)=g^{(2)}_3(z)+g^{(3)}_3(z)+g^{(4)}_3(z)$ as follows:
\begin{align}
\label{WW-g3-tw2}
g^{(2)}_3(z)=&\int_z^1\frac{\d y}{y}\Big\{\frac{g_1(y)}{2}
+\ln\Big(\frac{z}{y}\Big)g_1(y)\Big\},\\
g^{(3)}_3(z)=&-2\int_z^1 \frac{\d y}{y}\Big(g_1+g_2\Big)(y)
	      -2\int_z^1 \frac{\d y}{y^2}\int_y^1\d u\, g_2(u)\nonumber\\
	     &-2\int_z^1 \frac{\d y}{y}\ln\Big(\frac{z}{y}\Big)\big(g_1+g_2\big)(y)
	      -2\int_z^1 \frac{\d y}{y^2}\ln\Big(\frac{z}{y}\Big)
	      \int_y^1 \d u\, g_2(u),\\
g^{(4)}_3(z)=&\,g_3(z)+\frac{1}{2}
		\int_z^1 \frac{\d y}{y}\Big(3g_1+4g_2\Big)(y)
	      +2\int_z^1 \frac{\d y}{y^2}\int_y^1\d u\, g_2(u)\nonumber\\
	     &+\int_z^1 \frac{\d y}{y}\ln\Big(\frac{z}{y}\Big)
	     \big(g_1+2g_2\big)(y)
	      +2\int_z^1 \frac{\d y}{y^2}\ln\Big(\frac{z}{y}\Big)
	      \int_y^1\d u\, g_2(u).
\end{align}
Due to the fact that $g_3(z)$ contains twist-2, twist-3 as well as twist-4
we have obtained three integral relations. For example, Eq.~(\ref{WW-g3-tw2})
demonstrates that the twist-2 part $g^{(2)}_3(z)$ can be expressed in terms
of the twist-2 function $g_1(z)$.
The relations for the moments are:
\begin{align}
\label{WW-g3-tw2-n}
g^{(2)}_{3 n}=&\frac{1}{2(n+1)}\, g_{1 n}-\frac{1}{(n+1)^2}\,g_{1 n},\\
g^{(3)}_{3 n}=&-\frac{2}{n+1}\Big(g_{1 n}+g_{2 n}\Big)
	       -\frac{2}{(n+1)n}\, g_{2n}
	     +\frac{2}{(n+1)^2}\Big(g_{1 n}+g_{2 n}\Big)
	      +\frac{2}{(n+1)^2 n}\, g_{2 n},\qquad n>0\\
g^{(4)}_{3 n}=&\,g_{3 n}
		+\frac{1}{2(n+1)}\Big(3g_{1 n}+4g_{2 n}\Big)
	       +\frac{2}{(n+1)n}\, g_{2n}
	     -\frac{1}{(n+1)^2}\Big(g_{1 n}+2g_{2 n}\Big)
	      -\frac{2}{(n+1)^2 n}\, g_{2 n},\qquad n>1.
\end{align}
For $n=0,1$ in (\ref{WW-g3-tw2-n}) we find the sum rules
\begin{align}
\int_0^1\d z\ g_3(z) = - \frac{1}{2} \int_0^1 \d z\ g_1(z),
\qquad
\int_0^1\d z\, z g^{(2)}_3(z)=0,
\end{align}
as well as 
\begin{align}
\int_0^1 \d z\ z\ g_3^{(3)}(z) =
-\frac{1}{2} \int \d z\ z (g_1 + 2 g_2)(z)
\end{align}

Substituting (\ref{rel-f1}) into (\ref{rel-f4}), we get
the integral relations for the twist-2 and twist-4 part of 
$f_4(z)=f^{(2)}_4(z)+f^{(4)}_4(z)$
\begin{align}
f^{(2)}_4(z)&=\frac{1}{2} \int_z^1 \frac{\d y}{y}f_1(y)\\
f^{(4)}_4(z)&=f_4(z) -\frac{1}{2}\int_z^1 \frac{\d y}{y}f_1(y).
\end{align}
The corresponding relations for the moments read
\begin{align}
f^{(2)}_{4 n}&=\frac{1}{2(n+1)}\, f_{1 n},\\
f^{(4)}_{4 n}&=f_{4 n} -\frac{1}{2(n+1)}\,f_{1 n},\qquad n>0.
\end{align}

If we now substitute (\ref{rel-h1}) into (\ref{rel-hL}) and using
$h_L(z)=h_2(z)+h_1(z)/2$ and $h_2(z)=h^{(2)}_2(z)+h^{(3)}_2(z)$,
we derive
\begin{align}
\label{WW-JJ-tw2}
h^{(2)}_2(z)&=-\frac{h_1(z)}{2} + 2z\int_z^1 \frac{\d y}{y^2}h_1(y)\\
\label{WW-JJ-tw3}
h^{(3)}_2(z)&=h_2(z) + \frac{h_1(z)}{2}- 2z\int_z^1 \frac{\d y}{y^2}h_1(y),
\end{align}
where (\ref{WW-JJ-tw2}) is a twist-2 Wandzura-Wilczek--like relation which 
was obtained earlier by Jaffe and Ji~\cite{Jaf92} 
(see also~\cite{Kodaira99}). 
Obviously, (\ref{WW-JJ-tw3}) is the corresponding twist-3 relation.
For the moments we obtain
\begin{align}
\label{WW-JJ-tw2-n}
h^{(2)}_{2 n}&=-\frac{h_{1 n}}{2} + \frac{2}{n+2}\,h_{1 n},\\
\label{WW-JJ-tw3-n}
h^{(3)}_{2 n}&=h_{2 n} + \frac{h_{1 n}}{2}-\frac{2}{n+2}\, h_{1 n},\qquad n>0.
\end{align}
For $n=0,1,2$ in (\ref{WW-JJ-tw2-n}) we observe the following sum rules
\begin{align}
\int_0^1\d z\, h_2(z) = \frac{1}{2} \int_0^1\d z\ h_1(z),
\qquad
\int_0^1\d z\, z h^{(2)}_2(z) = \frac{1}{6} \int_0^1\d z\, z h_2(z),
\qquad
\int_0^1\d z\, z^2 h^{(2)}_2(z)=0.
\end{align}

Furthermore, 
using the formulas (\ref{rel-h1}), (\ref{rel-hL}) (\ref{rel-h3}) and
(\ref{rel-H3}), we obtain the integral relations for the structure function
$h_3(z)=h^{(2)}_3(z)+h^{(3)}_3(z)+h^{(4)}_3(z)$ as follows:
\begin{align}
\label{WW-h3-tw2}
h^{(2)}_3(z)=&\int_z^1\frac{\d y}{y}h_1(y)-z\int_z^1\frac{\d y}{y^2}h_1(y)\\
h^{(3)}_3(z)=&z\int_z^1 \frac{\d y}{y^2}\Big(h_2+\frac{h_1}{2}\Big)(y)
	      +2z\int_z^1 \frac{\d y}{y^3}\int_y^1\d u\Big(h_2-\frac{h_1}{2}\Big)(u)\\
h^{(4)}_3(z)=&\,h_3(z)-\int_z^1\frac{\d y}{y} h_1(y)
	     -z\int_z^1 \frac{\d y}{y^2}\Big(h_2-\frac{h_1}{2}\Big)(y)
	     -2z\int_z^1 \frac{\d y}{y^3}\int_y^1\d u\Big(h_2-\frac{h_1}{2}\Big)(u),
\end{align}
and the relations for the corresponding moments
\begin{align}
\label{WW-h3-tw2-n}
h^{(2)}_{3 n}=&\frac{1}{n+1}\,h_{1 n}-\frac{1}{n+2}\,h_{1 n},\\
h^{(3)}_{3 n}=&\frac{1}{n+2}\Big(h_{2 n}+\frac{h_{1 n}}{2}\Big)
	      +\frac{2}{(n+2)n}\Big(h_{2 n}-\frac{h_{1 n}}{2}\Big),\qquad n>0\\
h^{(4)}_{3 n}=&\,h_{3 n}-\frac{1}{n+1}\,h_{1 n}
		-\frac{1}{n+2}\Big(h_{2 n}-\frac{h_{1 n}}{2}\Big)
	      -\frac{2}{(n+2)n}\Big(h_{2 n}-\frac{h_{1 n}}{2}\Big),\qquad n>1.
\end{align}
Also here we obtain some useful sum rule from (\ref{WW-h3-tw2-n}) in the
case $n=0$, namely,
\begin{align}
\int_0^1\d z\, h_3(z) = \frac{1}{2}\int_0^1 \d z\, h_1(z).
\end{align}
Let us note that Jaffe and Ji ignored in their investigations~\cite{Jaf92} the function $h_3(z)$ 
because they claimed that it is twist-4 and can be neglected.
However, this assumption is only true in the case of the vector operator 
$M_{5\alpha}(\kappa_1\lcx,\kappa_2\lcx)$ but not in the
case of the tensor operator $M_{5[\alpha\beta]}(\kappa_1\lcx,\kappa_2\lcx)$.
As we have seen in Eq. (\ref{rel-h3}),
then the function $h_3(z)$ contains twist-2, twist-3 as well as twist-4 parts.
Moreover, the twist-2 part $h_3^{(2)}(z)$ is given in terms of the twist-2
function $h_1(z)$ in (\ref{WW-h3-tw2}). Therefore, the structure function 
$h_3(z)$ is experimentally relevant also at the twist-2 and twist-3 level and 
cannot be ignored in these cases.

\section{Conclusions}
Using the notion of geometric twist, we discussed the calculation 
of the forward matrix elements for those nonlocal LC-operators which 
correspond to the independent tensor structures in polarized 
nucleon matrix elements
of the type $\langle PS| \overline\psi(\kappa_1 \lcx) \Gamma 
U(\kappa_1 \lcx, \kappa_2 \lcx) \psi(\kappa_2\lcx)|PS \rangle$.
We have found nine independent forward distribution functions with well
defined twist $\tau$ which, for twist $\tau \geq 3$, differ from the
conventional ones. From the field theoretical point of view this Lorentz 
invariant classification is the most appropriate frame of introducing
distribution functions since the separation of different (geometric) twist 
is unique and independent from the special kinematics of the process.
Only the operators of definite geometric twist will have the correct mixing 
behaviour under renormalization.
A very useful property of the nonlocal (and local) LC operators of definite
twist $\tau$ is that they are obtained from the original, i.e.,~undecomposed
LC--operators by the application of corresponding projection operators.
An essential result of our calculations is the relation between the new
parton distribution functions and those given by Jaffe and Ji~\cite{Jaf92}.
In addition we have also given the moments of all the distribution
functions. 
These relations demonstrate
the interrelations between the different twist definitions.
One principal result of this paper  are Wandzura-Wilczek--like relations
between the dynamical twist distributions which we have obtained by means of 
our interrelation rules in natural manner. 
An advantage in our approach was that we used operators with geometric twist 
which allowed us to reveal the
Wandzura-Wilczek--like relations between the dynamical twist distribution
functions.
The same procedure may be applied to the twist decomposition
of the meson wave functions \cite{L00,L00b}.
Finally, in Appendix \ref{inner}, we have also written
the nonlocal (and local) LC--operators by using the inner derivatives
on the light--cone.

\acknowledgments
The authors are grateful to J. Bl\"umlein, D.~Robaschik and S.~Neumeier 
for useful discussions.
They also grateful acknowledge stimulating discussions during the Leiden
workshop on Polarized and Unpolarized Deep Inelastic Lepton--Hadron 
Scattering. M.L. acknowledges P.~Ball and A.V.~Radyushkin for 
stimulating discussions, furthermore to V.K.~Dobrev for correspondence.
Additionally, M.L. would like to express his gratitude to the 
Graduate College ``Quantum field theory'' at Center for Theoretical Sciences of
Leipzig University for financial support. 
We would like to thank the anonymous referee for some helpful comments.

\begin{appendix}
\section{Twist operators and the interior derivative on the light-cone}
\label{inner}
\renewcommand{\theequation}{\thesection.\arabic{equation}}
\setcounter{equation}{0}

In this Appendix we like to rewrite the LC--operators of definite twist
$\tau$ and, in addition, we formulate the property of tracelessness of 
such LC--operators by using internal derivatives on the light--cone.

Let us now introduce the notion of {\it interior differential operators} 
on the light-cone $K_{4}=\{\lcx\in{\Bbb M}^{4},\lcx^2=0\}$ which,
for the first time, has been used in order to characterize (irreducible)
symmetric tensor representations of $SO(4)$ on the cone and their 
graded algebra $P(K^n_{4})=\oplus_{n=0}^\infty K^n_{4}$, where 
$K^n_{4}$ is the space of homogeneous polynomials $T_n (\lcx)$ of degree $n$ 
on the cone~\cite{Bargmann,Dobrev77}:

A differential operator $Q$ is said to be an {\em interior} differential 
operator iff 
\begin{equation}
Q\big(x^2 T_n (x)\big){\big|}_{x^2\equiv \lcx^2=0}=0.
\end{equation}
For example, the generators of dilation $X=1+\lcx\lcd$ 
and rotations $X_{\mu\nu}=\lcx_\nu\lcd_\mu-\lcx_\mu\lcd_\nu$
are first order interior differential operators on the light-cone.
Obviously, they leave the space $K^n_4$ {\em invariant}. 

A further 
interior differential operator $\d_\mu$ of second order may be introduced
by the following requirements:\\
(a) it should be a {\em lowering operator}, i.e.~mapping a homogeneous 
polynomial of degree $n$ (of $\lcx$) into a homogeneous polynomial of 
degree $n-1$, $K_{4}^{n}\rightarrow K_{4}^{n-1}$;\\
(b) it should behave as a vector under rotations,
\begin{equation}
\label{lie_1}
[X_{\mu\nu},\d_\lambda]=\delta_{\mu\lambda}\d_\nu-\delta_{\nu\lambda}\d_\mu;
\end{equation}
(c) it should be the lowest order differential operator satisfying 
(a) and (b).\\
Choosing the normalization of this interior derivative 
so that $2\d_\mu$ is the generator
of {\it special conformal transformations} of 
{\it massless} 0-helicity representations of the conformal Lie algebra 
${\frak{so}}(4,2)$, the interior derivative is given as
\begin{align}
\d_\alpha f(\lcx)\equiv
\Big\{\big(1+x\pd\big)\pd_\alpha
-\hbox{\large$\frac{1}{2}$}
x_\alpha\square\Big\}f(x)\big|_{x=\lcx},
\end{align}
with the following properties
\begin{align}
\d^2 = 0\, ,\quad
[\d_\alpha,\d_\beta]=0
\quad\text{and}\quad
\d_\alpha\lcx^2=\lcx^2\big(\d_\alpha+2\lcd_\alpha\big)\,.
\end{align}
Obviously, $\d_\mu$ is a second order interior differential operator
which together with $\lcx_\mu$, $X$ and $X_{\mu\nu}$ satisfies 
the conformal {\em Lie algebra} ${\frak so}(4,2)$ in $\lcx$-space;
especially there holds the commutator relation
\begin{equation}
\label{lie_3}
[\d_\mu,\lcx_\nu]=\delta_{\mu\nu} X+X_{\mu\nu}.
\end{equation}
In that terminology $\lcx$ is a 
{\em raising operator} which plays the role of ``momentum''
in the conformal algebra:
\begin{align}
\label{lie_2}
[\lcx_\mu,\lcx_\nu]=0,\quad & \lcx^2=0.
\end{align}

Now, we rewrite the nonlocal LC--operators (\ref{O2vec}) -- 
(\ref{M_tw3_vec}) with the help of the above interior derivatives:
\begin{align}
\label{O2vec_d}
O^{\mathrm{tw2}}_{\alpha}(\kappa_1\lcx,\kappa_2\lcx)
&=
-\int_{0}^{1} \d\lambda\, (\ln\lambda) \,\d_\alpha \lcx^\mu \,
O_\mu(\kappa_1\lambda \lcx, \kappa_2\lambda \lcx)
\\
\label{O3vec_d}
O^{\mathrm{tw3}}_{\alpha}
(\kappa_1\lcx,\kappa_2\lcx)
&=
- \int_{0}^{1}\d\lambda\,(\ln\lambda)\,
\big\{
X^2\delta_\alpha^\mu -\d_\alpha \lcx^\mu-\lcx_\alpha\d^\mu
\big\}
 O_\mu(\kappa_1\lambda \lcx, \kappa_2\lambda \lcx)
\\
\label{O4vec_d}
O^{\mathrm{tw4}}_{\alpha}
(\kappa_1\lcx,\kappa_2\lcx)
&=-
\int_{0}^{1}\d\lambda\, (\ln\lambda)\,\lcx_\alpha\d^\mu\,
 O_\mu(\kappa_1\lambda \lcx, \kappa_2\lambda \lcx)
\\
\label{M_tw2_ten_d}
M^{\mathrm{tw2}}_{[\alpha\beta]}(\kappa_1\lcx,\kappa_2\lcx)
&=
-2\int_{0}^{1}\d\lambda\big(1-\lambda\big)
\Big\{
\d_{[\alpha} \delta_{\beta]}^{[\mu}\lcx^{\nu]}
+X_{[\alpha\beta]}X^{[\mu\nu]}
\Big\}
M_{[\mu\nu]}(\kappa_1\lambda \lcx, \kappa_2\lambda \lcx)
\\
\label{M_tw3_ten_d}
M^{\rm tw3}_{[\alpha\beta]}(\kappa_1\lcx,\kappa_2\lcx)
&=\int_{0}^{1}{\d\lambda}\,\hbox{\Large$\frac{1-\lambda^2}{\lambda}$}
\Big\{
\Big((\lcx\d) \delta_{[\alpha}^{[\mu}\delta_{\beta]}^{\nu]} 
- 2\delta_{[\alpha}^{[\mu}\lcx^{\nu]}\d_{\beta]}\Big) \nonumber\\
&\qquad\qquad
+2 \Big(
\lcx_{[\alpha}\delta_{\beta]}^{[\mu}\d^{\nu]}
-X_{[\alpha\beta]}X^{[\mu\nu]}\Big)
\Big\}
M_{[\mu\nu]}(\kappa_1\lambda \lcx, \kappa_2\lambda \lcx)
\\
&\ \ \ -\int_{0}^{1}\d\lambda\,
\hbox{\Large$\frac{1-\lambda^2}{\lambda}$}
X_{[\alpha\beta]}X^{[\mu\nu]}
M_{[\mu\nu]}(\kappa_1\lambda \lcx, \kappa_2\lambda \lcx)
\label{M_tw3_ten_d1}
\\
\label{M_tw4_ten_d}
M^{\rm tw4}_{[\alpha\beta]}(\kappa_1\lcx,\kappa_2\lcx)
&=2\int_0^1\d\lambda\,
\hbox{\Large$\frac{1-\lambda}{\lambda}$}
\Big\{
X_{[\alpha\beta]}X^{[\mu\nu]}
-\lcx_{[\alpha}\delta_{\beta]}^{[\mu} \d^{\nu]}
\Big\}
M_{[\mu\nu]}(\kappa_1\lambda \lcx, \kappa_2\lambda \lcx)
\\
\label{M_tw2_vec_d}
M^{\mathrm{tw2}}_{\alpha}(\kappa_1\lcx,\kappa_2\lcx)
&=-\int_0^1\d\lambda\, \lambda\ln\lambda\,
\Big\{ 
X^2\delta_\alpha^\mu-\lcx_\alpha \d^\mu
\Big\}
M_{\mu}(\kappa_1\lambda \lcx, \kappa_2\lambda \lcx)
\\
\label{M_tw3_vec_d}
M^{\mathrm{tw3}}_{\alpha}(\kappa_1\lcx,\kappa_2\lcx)
&=
-\int_0^1\d\lambda\,\lambda\ln\lambda\,\lcx_\alpha\d^\mu\,
M_\mu(\kappa_1\lambda \lcx, \kappa_2\lambda \lcx)\ .
\end{align}
Let us mention that the actual tensor and vector LC-operators satisfy 
conditions of tracelessness on the light-cone which can be formulated by means of the 
interior derivative $\d_\mu$~\cite{Dobrev82}. 
For example, these conditions simply read (see~\cite{gl99c,gl00a}):
\begin{align}
\d^\alpha O^{\mathrm{\tau=2,3}}_{\alpha}(\kappa_1\lcx,\kappa_2\lcx)=0,
\qquad
\d^\alpha M^{\mathrm{\tau=2}}_{[\alpha\beta]}(\kappa_1\lcx,\kappa_2\lcx)=0,
\qquad
\d^\alpha M^{\mathrm{\tau=2}}_{\alpha}(\kappa_1\lcx,\kappa_2\lcx)=0.
\end{align}
Scalar operators on the light-cone are traceless by construction since they
correspond to totally symmetrized tensors.

These relations may be proven very easily by using the above mentioned 
properties of inner derivatives. Also the proof of the projection properties 
is straightforward but tedious.

Let us also rewrite the local twist operators (\ref{O2vec_l}) --
(\ref{M_tw3_vec_l}) by means of inner derivatives: 
\begin{align}
\label{O2vec_l_d}
O^{\mathrm{tw2}}_{\alpha n}(\lcx)
&=
\hbox{\Large$\frac{1}{(n+1)^2}$}\,
\d_\alpha O_{n+1}(\lcx)
\\
\label{O3vec_l_d}
O^{\mathrm{tw3}}_{\alpha n}(\lcx)
&=
\hbox{\Large$\frac{1}{(n+1)n}$}
\Big(n^2\delta_\alpha^\mu-\lcx^\mu\d_\alpha\
-\hbox{\Large$\frac{n-1}{n+1}$}
\lcx_\alpha\d^\mu\Big) O_{\mu n}(\lcx)
\\
\label{O4vec_l_d}
O^{\mathrm{tw4}}_{\alpha n}(\lcx)
&=\hbox{\Large$\frac{1}{(n+1)^2}$}\,
\lcx_\alpha\d^\mu O_{\mu n}(\lcx)
\\
\label{M_tw2_ten_l_d}
M^{\mathrm{tw2}}_{[\alpha\beta]n}(\lcx)
&=\hbox{\Large$\frac{2}{(n+2)(n+1)}$}
 \Big\{\d_{[\beta}\delta_{\alpha]}^\mu
-\hbox{\Large$\frac{1}{n+2}$}\, \lcx_{[\alpha}\lcd_{\beta]}\d^\mu\Big\}
M_{\mu\, n+1}(\lcx)
\\
\label{M_tw3_ten_l_d}
M^{\rm tw3}_{[\alpha\beta]n}(\lcx)
&=
\hbox{\Large$\frac{1}{(n+2)n}$}
\Big\{\delta_{[\alpha}^\mu\delta_{\beta]}^\nu n^2
- 2\lcx^\nu\delta_{[\alpha}^\mu\d_{\beta]} 
+\hbox{\Large$\frac{2}{n}$}\, \lcx_{[\alpha}
\Big(\delta_{\beta]}^\mu(n-1)-\lcx^{\mu}\lcd_{\beta]}\Big)\d^\nu
\Big\}M_{[\mu\nu] n}(\lcx)
\nonumber\\
&\ \ \ +\hbox{\Large$\frac{2}{(n+2)^2n}$}\,
\lcx_{[\alpha}\lcd_{\beta]}\d^\mu M_{\mu\, n+1}(\lcx)
\\
\label{M_tw4_ten_l_d}
M^{\rm tw4}_{[\alpha\beta]n}(\lcx)
&=
\hbox{\Large$\frac{2}{(n+1)n}$}
\Big\{\lcx_{[\alpha}\lcd_{\beta]}\lcx^{[\mu}\lcd^{\nu]}
-\lcx_{[\alpha}\delta_{\beta]}^{[\mu} \d^{\nu]}\Big\}
M_{[\mu\nu] n}( \lcx)
\\
\label{M_tw2_vec_l_d}
M^{\mathrm{tw2}}_{\alpha\, n+1}(\lcx)
&=
M_{\alpha\, n+1}(\lcx)
-\hbox{\Large$\frac{1}{(n+2)^2}$}\,
\lcx_\alpha \d^\mu
M_{\mu\, n+1}(\lcx)
\\
\label{M_tw3_vec_l_d}
M^{\mathrm{tw3}}_{\alpha\, n+1}(\lcx)
&=
\hbox{\Large$\frac{1}{(n+2)^2}$}\,
\lcx_\alpha\d^\mu
M_{\mu\, n+1}( \lcx).
\end{align}

Now we point out some technical details.
The local operators of definite twist are written in such a manner that
their relation to the various symmetry classes by which they are
determined becomes more obvious (for details see~\cite{glr99b,gl00a}).
The twist--4 operator (\ref{O4vec_l_d}) is built up by those parts
of the operators (\ref{O2vec_l_d}) and (\ref{O3vec_l_d}) which are
related to the Young pattern $(n-1)$, i.e., being contained in the trace 
terms. The twist--3 operator (\ref{M_tw3_ten_l_d}) consists of 
two independent operators, namely a twist--3 operator
corresponding to the traces of the Young 
pattern $(n,2)$ which obey the totally symmetric pattern $(n)$,
and another operator which is constructed by means of the
Young pattern $(n,1,1)$. Both terms have been separately written,
cf. also Eqs.~(\ref{M_tw3_ten_d}) and (\ref{M_tw3_ten_d1}).
The twist--4 operator (\ref{M_tw4_ten_l_d}) contains 
the $(n-1,1)$ part of the traces of the Young 
patterns $(n,2)$ and $(n,1,1)$. 

Let us discuss the conditions of tracelessness for the local light--cone 
operators in more detail. 
For the leading twist operators these conditions are
\begin{align}
\d^\alpha O^{\tau=2,3}_{\alpha n}(\lcx)=0,
\qquad
\d^\alpha M^{\tau=2}_{[\alpha\beta]n}(\lcx)=0,
\qquad
\d^\alpha M^{\tau=2}_{\alpha n+1}(\lcx)=0.
\end{align}
Properly speaking, the higher twist parts of the vector operators 
$O^{\mathrm{tw4}}_{\alpha\, n+1}(\lcx)$ and 
$M^{\mathrm{tw3}}_{\alpha\, n+1}(\lcx)$ are 
$\d^\mu O_{\mu\, n}( \lcx)$ and $\d^\mu M_{\mu\, n+1}( \lcx)$, respectively.
In fact, these operators are scalar operators being traceless 
on the light-cone.
The case of the higher twist tensor operators is more complicated.
The condition of tracelessness for the twist-3 tensor operator 
corresponding to the pattern $(n,1,1)$, i.e. the first term 
in~(\ref{M_tw3_ten_l_d}), reads
\begin{align}
\d^\alpha M^{\tau=3}_{[\alpha\beta]n}(\lcx)=0.
\end{align}
Actually, the other term is a twist-3 vector
operator $\widetilde{M}^{\rm tw3}_{\beta n-1}(\lcx)$
multiplied by $\lcx_\alpha$,
\begin{align}
M^{\rm tw3}_{[\alpha\beta]n}(\lcx)
=
\hbox{\Large$\frac{2}{(n+2)^2 n^2}$}\,
\lcx_{[\alpha}\widetilde{M}^{\rm tw3}_{\beta] n-1}(\lcx),\quad
\widetilde{M}^{\rm tw3}_{\beta\, n-1}(\lcx)=\d_{\beta}\d^\mu M_{\mu\, n+1}(\lcx),
\end{align}
and also the twist-4 tensor operator is given analogously by a 
vector operator $\widetilde{M}^{\rm tw4}_{\beta n}(\lcx)$,
\begin{align}
M^{\rm tw4}_{[\alpha\beta]n}(\lcx)
=
\hbox{\Large$\frac{2}{(n+1)n}$}\,
\lcx_{[\alpha}\widetilde{M}^{\rm tw4}_{\beta] n-1}(\lcx),\quad
\widetilde{M}^{\rm tw4}_{\beta\, n-1}(\lcx)=
\Big\{\hbox{\Large$\frac{1}{n}$}\d_{\beta}\lcx^{[\mu}\lcd^{\nu]}
-\delta_{\beta}^{[\mu} \d^{\nu]}\Big\}
M_{[\mu\nu] n}( \lcx).
\end{align}
The conditions of tracelessness for these vector operators read
\begin{align}
\d^\beta \widetilde{M}^{\rm tw3}_{\beta n-1}(\lcx)=0, \qquad
\d^\beta \widetilde{M}^{\rm tw4}_{\beta n-1}(\lcx)=0.
\end{align}
It is clear that after multiplication with $\lcx_\alpha$ and
antisymmetrization with respect to $\alpha$ and
$\beta$ the trace terms of these vector operators are cancelled.

Each series of local operators (\ref{O2vec_l_d}) -- (\ref{M_tw3_vec_l_d})
may be re-expressed by one nonlocal operator by using formulas like
$ {\kappa^{n}}/{(n+1)^2}
=- \int_0^1 \d\lambda (\ln\lambda)(\kappa\lambda)^n $
as easily can be seen in the case of twist--2 vector operators.
The other fractions are somewhat more involved; factors of $n$ in
the nominator are to be expressed by $(\lcx\lcd)$.

The expressions for the twist--$\tau$ operators in terms of 
internal derivatives is not unique because, by using formulas like
$\lcx^{[\mu}\d^{\nu]} = (\lcx\lcd)\lcx^{[\mu}\lcd^{\nu]}$ and
performing partial $\lambda$--integrations, they may be reformulated.
This has been done for the nonlocal operators
(\ref{O2vec_d}) -- (\ref{M_tw3_vec_d}) at various places in order
to obtain a form where their projection properties can be proven
most easily by using the commutation relations between the 
generators $\lcx_\mu, \d_\mu, X$ and $X_{\mu\nu}$. Obviously,
some of these reformulations are true only in the case of the antisymmetric
tensor operator whereas the corresponding local expressions hold more
generally.

Finally, let us mention that Dobrev {\it et al.}~\cite{Dobrev77,Cragie} used the 
interior derivative $\d_\mu$ for the construction of local symmetric 
(traceless)
light-cone operators which carry an elementary (irreducible) representation
of the conformal group $SO(4,2)$. 

\section{Transformation properties of nonlocal LC operators}
\setcounter{equation}{0}

Because the notion of geometric twist is related to the framework of the 
group 
$SO(3,1)\otimes {\Bbb R}_+$, where ${\Bbb R}_+$ is the group of dilations, 
it is useful to give the transformation rules of the (unrenormalized) 
nonlocal operators
with respect to this group. We induce the transformation rules of the 
nonlocal operators from the transformation properties of local
fields which are given in~\cite{Dobrev76,Dobrev77}. 
Let $\OO^\tau_{\alpha n}(y,\lcx)$ be the local components -- being 
obtained by a formal Taylor expansion around the point $y$ --
of any nonlocal operator 
$\OO^\tau_{\alpha}(y-\lcx,y+\lcx)$ (with $\kappa_1=-1$, $\kappa_2=1$) of
geometric twist $\tau$.
These local operators
$\OO^\tau_{\alpha n}(y,\lcx)$ are infinitely differentiable with respect to
$y\in{\Bbb R}^4$ and homogeneous 
polynomials of degree $n$ in the light-like vector $\lcx$.

The transformation properties of these local operators 
are~~\cite{Dobrev76,Dobrev77}\\
a) Poincar{\'e} transformations:
\begin{align}
\hat U(a,\Lambda) \OO^\tau_{\alpha n}(y,\lcx)\hat U^{-1}(a,\Lambda)
= V_{\alpha\beta}(\Lambda)\OO^\tau_{\beta n}(\Lambda^{-1}(y-a),\Lambda^{-1}\lcx),
\quad a\in{\Bbb R}^4, \Lambda\in SO(3,1),
\end{align}
where $V(\Lambda)$ is a finite dimensional 
representation of the Lorentz group. 
Obviously, the translation $\hat U(a,0)=\e^{-\ii a\cdot\Hat P}$ with
$\Hat P_\mu=\ii\frac{\pd}{\pd y^\mu}$ just shifts the expansion point
$y$ by the vector $a$. If we would have chosen $a=\lcx$, the twist of these
operators would be not invariant under this translation, 
because the differential operators for definite twist do not commute
with $\lcx$.
\\
b) Dilations:
\begin{align}
\hat U(\rho) \OO^\tau_{\alpha n}(y,\lcx)\hat U^{-1}(\rho)
= \rho^d \OO^\tau_{\beta n}(\rho y,\lcx),
\quad \rho > 0,
\end{align}
where $d$ is the (scale) dimension of the local operator (in mass units).

The transformation properties for the nonlocal operators are\\
a) Poincar{\'e} transformations:
\begin{align}
\hat U(a,\Lambda) \OO^\tau_{\alpha}(y-\lcx,y+\lcx)\hat U^{-1}(a,\Lambda)
= V_{\alpha\beta}(\Lambda)\OO^\tau_{\beta}(\Lambda^{-1}(y-a-\lcx),\Lambda^{-1}(y-a+\lcx)),
\end{align}
b) Dilations:
\begin{align}
\hat U(\rho) \OO^\tau_{\alpha}(y-\lcx,y+\lcx)\hat U^{-1}(\rho)
= \rho^d \OO^\tau_{\alpha }(\rho (y-\lcx),\rho (y+\lcx)).
\end{align}
Let us point out that in the case nonlocal operators, i.e., for an
infinite series of local ones, we deal 
with an infinite-dimensional representation of the dilation and 
the Lorentz group.         
\end{appendix}

\end{document}